\documentclass[letterpaper]{article}
\usepackage[preprint]{aaai2027}
\usepackage[hyphens]{url}
\usepackage{graphicx}
\usepackage{natbib}
\usepackage{caption}
\usepackage{booktabs}
\usepackage{amsmath,amssymb}
\usepackage{multirow}

\title{The Uncontrolled Variable: Vision--Language Refusal Is Conditioned on the Image-Attachment Interface, and Not Robust to Irrelevant Image Properties}
\author{
    Haoyu Zhang,
    Yi Feng,
    Hanwen Liu,
    Shibo Zheng,
    Zhuoxi Wang,
    Yang Chen,
    Haowen Xu,
    Xiangchen Guan,
    Mohammad Zandsalimy,
    Shanu Sushmita
}
\affiliations{}

\begin{document}
\maketitle

\begin{abstract}
A safety-aligned model is supposed to decide whether to answer on the basis of what is being asked. We show that aligned vision--language models also condition refusal on a property of a request's \emph{form}: whether an image is attached, holding everything the request asks fixed. Attaching a \emph{blank canvas}, an image that cannot be read, cannot relate to the request, and is byte-identical across every prompt in its condition, shifts benign refusal by tens of points. Every measurement here is taken with \emph{no defense in the loop}; the only manipulation anywhere is attachment. The shift is not blanket caution but a threshold shift: genuinely neutral instructions are almost unaffected ($\leq2$ percentage points on three of four hosted models) while borderline-benign prompts move $+23$ to $+51$ points, so the cost falls on sensitivity-adjacent traffic, meaning benign questions about privacy, self-harm, violence and illegal activity. There is a benign reading of such a threshold, namely that attachment correlates with risk in real traffic, and we take it seriously; a black-box study cannot measure that correlation and we do not claim to. What it can test is whether the response to attachment is robust to variation carrying no information about the request, and on four independent measurements it is not. It varies with canvas colour and pixel count. Its sign inverts across checkpoints. It survives an explicit instruction to disregard the image. And on one model it fires on a bare assertion that an attachment exists, with nothing attached and the modality word contributing none of it. We then localise the effect as far as black-box access allows: it is not a serving-stack artifact, one checkpoint reached through a managed host and through our own vLLM agreeing within a few points on every arm; it is not a property of vision--language models as such, three open-weight models showing nothing; it is a property of particular aligned checkpoints, and we exhibit an open one that reproduces the full hosted-scale effect with no moderation layer anywhere in the path. Finally we price it. On a matched harmful set the same canvas does lower attack success, so the cue buys something. But the charge is decoupled from the purchase: the checkpoint with the least harmful headroom we measure, completing only $2\%$ of plain harmful requests, still pays the benign cost in full, and across our models the harmful-side denominator falls as alignment improves while the benign cost does not track it down. The better a model already is, the more of this cue is cost without a matching purchase. Image presence is not a conservative default that a deployer chose and priced. It is an uncontrolled variable.
\end{abstract}

\section{Introduction}
\label{sec:intro}
A safety-aligned model is supposed to decide whether to answer on the basis of what is being asked. Vision--language models (VLMs) are deployed behind an interface where the user may or may not attach an image, and whether they do is not a fact about the request.

\paragraph{Why it goes unmeasured.}
Safety benchmarks vary what a request asks, and multimodal ones vary what the attached image \emph{contains}. Neither design holds content fixed while varying whether an image is attached at all, so the presence axis is not merely unmeasured: a threshold shift keyed on it is attributed to whatever content varied alongside it.

\paragraph{The cost is not spread evenly, which is what makes it an alignment cost.}
Across a three-rung sensitivity ladder, genuinely neutral instructions are almost unaffected ($\leq2$ percentage points on three of four hosted models) while borderline-benign prompts move $+23$ to $+51$ points ($p\leq1.5\times10^{-6}$; Table~\ref{tab:ladder}). The models are not becoming uniformly more cautious.

\paragraph{The claim, and what would refute it.}
There is a benign reading of a threshold that moves with attachment, and we take it seriously: users plausibly do attach images more often in sensitive settings, so conditioning on attachment could be rational inference about risk. A black-box study of shipped models cannot measure that correlation, and we do not claim to. What it \emph{can} test is whether the response to attachment is robust to variation carrying no information about the request. To be concrete about what would count as a legitimate attachment prior: a policy conditioning on \emph{whether} an image is attached should be invariant to \emph{which} uninformative image it is, because no such image carries information the prior could use.

\paragraph{Where the effect lives.}
The natural deflationary reading is that this is a vendor artifact, a moderation layer wrapped around a hosted endpoint, which would make it someone else's implementation detail rather than an alignment property. We rule that out in three steps. \emph{It is not the serving stack}: \texttt{gemma-3-12b-it} is an open checkpoint served both by a managed commercial host and by our own vLLM, and the identical manipulation gives route differences bounded inside $\pm10$ points on every arm and $\pm5$ points on both blank arms, with the same presence effect appearing down each. \emph{It is not VLMs as such}: on three open-weight models served with no moderation layer the benign cost is absent. \emph{It is a property of particular aligned checkpoints}, and we exhibit an open one. \texttt{qwen3-vl-8b}, self-served under vLLM with nothing wrapped around it, shows $+32$, $+28$ and $+29$ points on the borderline rung across three independently collected jobs: an effect the size of the frontier hosted models, on eight billion downloadable parameters. Finding it required looking, and a coarse ``alignment tier'' label does not predict it.

\paragraph{What it costs, what it buys, and why those are not the same question.}
A threshold shift should also catch more borderline-harmful requests, and it does: on OR-Bench's harmful split, the constructed counterpart of the borderline-benign rung, the same canvas lowers attack success by $5$ to $18$ points. We report the two sides side by side rather than as a ratio, for a reason that is itself part of the finding.

\paragraph{Why this is an alignment question.}
We claim only a shortcut, since a black-box experiment cannot establish that image attachment is unassociated with safety policy acquired in post-training. What we can state is that the response to that form is not robust to variation carrying no information about the request. The consequences are concrete on both sides of the interface. A benign user asking a privacy or self-harm question pays a refusal for attaching a photograph. A model whose refusals are partly keyed on attachment carries a control surface that an evaluation of its harm judgement is not built to reveal, because such an evaluation varies the request and holds the interface fixed. And because an open checkpoint reproduces the full hosted-scale effect with no vendor filter anywhere in the path, at least one form of it is a property a deployer inherits with the weights rather than one they can host their way out of.

\paragraph{Scope, and what we do not claim.}
We measure with no defense anywhere in the pipeline, so no result here can be attributed to a guardrail's design or evaluation, and we make no claim about any. We do not claim these effects are universal: we describe four hosted models and seven open-weight checkpoints, and we report the models on which the effect is absent as prominently as those on which it is large. We do not claim image attachment is uninformative in deployment; our constant canvas establishes only that it carries no \emph{per-prompt} information within an arm, which is true by construction, and nothing here rests on the correlation being absent.

\paragraph{Contributions.}
(i) \textbf{A threshold shift keyed on request form, isolated by a control that carries no per-prompt information.} A three-rung sensitivity ladder isolating \emph{who} pays the benign cost, stratified across all ten benchmark categories, with no defense anywhere in the pipeline (\S\ref{sec:res-threshold}).
(ii) \textbf{The response does not behave like a risk policy, on four independent measurements, and the obvious deflations fail.} It varies with canvas colour and pixel count with an image attached in both arms; its sign inverts across checkpoints; it survives an explicit instruction to disregard the image; and it fires on a bare asserted attachment. Separately, it is not a lexical artifact (attachment moves refusal $+20$ points with the mention held constant and \emph{no} prompt moving the other way), not an input-length effect (text padded to the canvas's own token budget buys nothing), not a serving-stack artifact, and not a property of VLMs as such (\S\ref{sec:res-threshold}).
(iii) \textbf{The charge is decoupled from what it prevents, and the decoupling widens as models improve.} Both sides measured on matched populations and reported side by side rather than as a ratio; an open checkpoint at $2\%$ residual attack success paying $+29$ points of benign cost; a harmful-side denominator that falls across our model set while the benign cost does not; and a sign that inverts outright on two open checkpoints (\S\ref{sec:res-threshold}).

\section{Manipulation and Method}
\label{sec:method}

\paragraph{The attached image carries no per-prompt information.}
Within an arm, every prompt receives the \emph{byte-identical} file: one distinct image
hash per condition, verified before each run. An image that does not vary across prompts
cannot carry information about any individual prompt, so a per-prompt behavioural
difference between the no-image and image arms cannot be a response to what the image
says about that request. We claim only this. A fixed canvas may still be statistically
associated with multimodal safety policy acquired in post-training; a black-box
experiment cannot exclude that, and our conclusions are stated behaviourally throughout.

\paragraph{Pairing protocol.}
All arms in a comparison are generated from one canonical text chained off the same
source step, so the text channel is byte-identical across arms and the arms differ only
in the attached file (and, in \S\ref{sec:res-threshold}'s factorial, in the system
message). Arms are paired per prompt id and tested with the exact McNemar test. \textbf{Every contrast is collected inside a single job}, because these targets are not
deterministic even at temperature $0$ and no provider we use exposes a seed, so a
numerator and a denominator from different windows would carry run drift into the
contrast.

\paragraph{What is varied.}
Four families of arm appear here. \textbf{Presence}: no image versus a blank canvas, on
each rung of a sensitivity ladder.

\section{Experimental Setup}
\label{sec:setup}

\textbf{Hosted models:} \texttt{claude-sonnet-4-6}, \texttt{gpt-4o-mini}, \texttt{gemini-2.5-flash-lite}, \texttt{gemini-2.5-flash}. \textbf{Open-weight models}, served by us under vLLM with no moderation layer of any kind: \texttt{qwen2-vl-7b}, \texttt{qwen2.5-vl-7b}, \texttt{qwen3-vl-8b-instruct} (abbreviated \texttt{qwen3-vl-8b} throughout), \texttt{internvl3-8b}, \texttt{pixtral-12b}, \texttt{llava-1.5-7b}, \texttt{gemma-3-12b-it}: the last also reachable through a managed host (AWS Bedrock), which is what makes the serving-route control possible. \textbf{Benign prompt sets}, in increasing topical sensitivity: $100$ genuinely neutral instructions drawn from AlpacaEval~\citep{li2023alpacaeval}, spanning all five of its constituent subsets (rung~1); the ``seemingly toxic but actually benign'' split of OR-Bench~\cite{cui2025orbench} (rung~2, where the cost lands); and the topic-matched benign split of JailbreakBench~\cite{chao2024jailbreakbench} (rung~3). \textbf{ASR}: the fraction of $100$ harmful prompts whose response is judged harmful by the HarmBench classifier~\cite{mazeika2024harmbench}; \textbf{lower is safer}.

\section{Results}
\label{sec:results}
\label{sec:res-threshold}

The effect measured here is a property of the target models themselves, and we measure it with \emph{no defense in the loop at all}: the only manipulation is whether a blank canvas is attached. Because the request text is carried in full in the text channel of every arm, and the canvas is byte-identical across all $100$ prompts of its condition (exactly one distinct image hash per condition), the image channel carries literally zero per-prompt information; the arms are paired per prompt id and differ in image presence alone.

\paragraph{How to read the table tags.}
Every table in this paper and in the Supplementary Document opens with a bracketed tag naming the evidentiary role each table plays: \textsc{primary test} (a declared family of \S\ref{app:multiplicity}, control arms included), \textsc{replication} (a result re-measured on a further model or in a fresh pass), \textsc{model-selection scan} (a sweep from which a checkpoint was chosen for later tables), \textsc{exploratory follow-up} (single-checkpoint or direction-establishing only, not powered to fix a cause). A companion Supplementary Document carries the full statistical protocol, the multiplicity families, the reproducibility record, and the arms reported here by pointer; section and table numbers prefixed with \textbf{S} refer to it.

\subsection{The shift, and who pays it}
\label{sec:res-shift}

\paragraph{The cost side scales with topical sensitivity.}
We measure benign refusal on three rungs of increasing sensitivity (Table~\ref{tab:ladder}): genuinely neutral instructions, ``seemingly toxic but actually benign'' OR-Bench prompts, and the topic-matched JailbreakBench-benign set. Inflation is near-absent on the neutral rung ($\leq2$ points on three of four models) and large on the borderline rung ($+23$ to $+51$ points on three of four, all $p\leq1.5\times10^{-6}$).

\begin{table}[t]
\centering
\scriptsize
\setlength{\tabcolsep}{3pt}
\begin{tabular}{lccc}
\toprule
 & \multicolumn{3}{c}{benign refusal \%, text\,$\to$\,blank canvas} \\
\cmidrule(lr){2-4}
Model & neutral & borderline & topic-matched \\
\midrule
claude-sonnet-4-6      & 2\,$\to$\,2   & 12\,$\to$\,\textbf{63}$^{***}$ & 3\,$\to$\,36 \\
gpt-4o-mini            & 2\,$\to$\,2   & 12\,$\to$\,\textbf{46}$^{***}$ & 4\,$\to$\,19 \\
gemini-2.5-flash-lite  & 1\,$\to$\,14$^{***}$  & 11\,$\to$\,\textbf{34}$^{***}$ & 11\,$\to$\,45 \\
gemini-2.5-flash       & 0\,$\to$\,2   & 16\,$\to$\,13 & 11\,$\to$\,18 \\
\bottomrule
\end{tabular}
\caption{[\textsc{primary test}]~\textbf{The sensitivity ladder}: benign refusal rate (\%, lower is more useful) under \{no image, blank canvas\}, no defense, $100$ prompts per rung, paired per prompt id. $^{***}p\leq1.5\times10^{-6}$, exact McNemar. Significance is marked on the neutral and borderline rungs, the two columns entering the declared test families (\S\ref{app:multiplicity}); unmarked cells in those two columns are non-significant. The topic-matched column is reported for continuity and is not part of any tested family, so it carries no marks. Inflation is absent on neutral prompts and large on borderline-benign ones: a threshold shift rather than blanket caution (\S\ref{sec:res-threshold}). The topic-matched column is the JailbreakBench-benign set, shown for continuity.}
\label{tab:ladder}
\end{table}

\paragraph{The cost is spread across topics, and the first measurement understated it.}
The borderline rung above draws its $100$ prompts from the head of OR-Bench's hard
split, and that file is ordered by category, so those prompts are
\emph{deception} and \emph{harassment} only, two of the benchmark's ten categories. That is a sampling defect in the measurement carrying our central cost claim, and it
is ours rather than the benchmark's. We therefore re-ran the borderline rung on a
category-stratified sample: $30$ prompts drawn from each of the ten categories, $300$
in total, same manipulation, same rubric, both arms in one job per model
(Table~\ref{tab:strata} of the Supplementary Document).

\paragraph{It is presence, not that particular image.}
A blank canvas is one image, so the ladder alone cannot distinguish ``image presence'' from ``blank canvases specifically''. We repeat rung~2 with a second, entirely different request-independent image (the clip-art line drawing of Fig.~\ref{fig:decoys}(c), a $1189\times1418$ black-and-white illustration) chained off the same canonical text, all arms in one job (Table~\ref{tab:presence}).

\begin{table}[t]
\centering
\scriptsize
\setlength{\tabcolsep}{3pt}
\begin{tabular}{lcccc}
\toprule
 & \multicolumn{4}{c}{benign refusal \%, rung 2 (borderline)} \\
\cmidrule(lr){2-5}
Model & text & blank & line draw. & blank+``ignore'' \\
\midrule
claude-sonnet-4-6      & 10 & \textbf{64}$^{***}$ & \textbf{58}$^{***}$ & 52 \\
gpt-4o-mini            & 13 & \textbf{43}$^{***}$ & \textbf{44}$^{***}$ & 36 \\
gemini-2.5-flash-lite  & 11 & \textbf{34}$^{***}$ & \textbf{63}$^{***}$ & 28 \\
gemini-2.5-flash       & 15 & 14 & \textbf{25}$^{*}$ & \textbf{47}$^{***}$ \\
\bottomrule
\end{tabular}
\caption{[\textsc{primary test, control arm}]~\textbf{Presence controls}, all arms in one job, one canonical text, $100$ prompts each, no defense. Two request-independent image classes, a blank canvas ($1024\times768$, pure white) and the clip-art line drawing of Fig.~\ref{fig:decoys}(c) ($1189\times1418$, drawn content, no text), plus the cheapest mitigation (a neutral instruction to disregard the image). The blank canvas here is generated at the renderer's default size, and is therefore a different file from the $512\times512$ blank of Fig.~\ref{fig:decoys}(b); both are pure white and neither carries readable text. $^{***}p<10^{-5}$, $^{*}p<0.05$ against the text baseline, exact McNemar paired per prompt; unmarked cells in the two image columns are non-significant against that baseline. The instruction column is a mitigation arm whose contrast of interest is against the \emph{blank} arm rather than against text, and those three contrasts are reported in the text; the one mark in that column is \texttt{gemini-2.5-flash}, where the instruction is not a mitigation but the manipulation itself and the text baseline is therefore the right comparator. Both classes inflate refusal on three of four models; the instruction removes only $6$--$12$ points of a $23$--$54$-point shift, and on \texttt{gemini-2.5-flash} it \emph{adds} $32$ points: an asserted-attachment effect rather than an image effect, decomposed in Table~\ref{tab:placebo}. The \texttt{gemini-2.5-flash} line-drawing cell ($+10$ points, $p=0.021$) does \emph{not} replicate: a larger property sweep puts the same contrast at $+9$ points, $p=0.064$ (\S\ref{app:imgprops}). We treat that model as near-null.}
\label{tab:presence}
\end{table}

\paragraph{The response varies with properties that carry no risk information.}
A policy conditioning on \emph{whether} an image is attached should be invariant to \emph{which} uninformative image it is, because no such image carries information the policy could use. That invariance is directly testable, and it fails. We hold image presence fixed (an image is attached in \emph{both} arms of every contrast, so these are not presence-versus-absence comparisons) and move one property at a time, paired per prompt id with Newcombe intervals (Table~\ref{tab:imgvsimg}).

\begin{table*}[t]
\centering
\scriptsize
\setlength{\tabcolsep}{3pt}
\begin{tabular}{llcccc}
\toprule
contrast (image attached in \emph{both} arms) & model & $\Delta$ & disc. & $p$ & $95\%$ CI \\
\midrule
black vs white, $512^2$        & fl.-lite & $+22$ & 22/0 & $4.8{\times}10^{-7}$ & $[+13.7,+29.7]$ \\
content vs blank, $1024{\times}141$ & fl.-lite & $+30$ & 31/1 & $1.5{\times}10^{-8}$ & $[+20.2,+38.7]$ \\
content vs blank, $1024{\times}141$ & claude   & $+28$ & 30/2 & $2.5{\times}10^{-7}$ & $[+18.0,+37.0]$ \\
content vs blank, $1024{\times}141$ & 4o-mini  & $+1$  & \phantom{0}5/4 & $1.00$ & $[-4.8,+6.8]$ \\
\addlinespace[2pt]
blank $1536^2$ vs blank $256^2$ & qwen3-vl & $+16$ & 18/2 & $0.0004$ & $[+7.7,+24.9]$ \\
\bottomrule
\end{tabular}
\caption{[\textsc{primary test}]~\textbf{The response is not invariant to properties that carry no risk information.} Each row compares two \emph{attached-image} arms against each other rather than against the text baseline, so image presence is held fixed and only the named property moves; size is matched within every row except the last, which varies size alone at fixed content. Every row is a paired exact McNemar contrast computed between the two image arms on matched prompt ids, with Newcombe intervals; $\Delta$ is the larger-property arm minus the smaller. The first four rows are the three hosted models, $n=100$ per arm; the last is the open checkpoint \texttt{qwen3-vl-8b}, also $n=100$. The paired tests resolve what raw rate differences could not: the \texttt{claude} content premium is established ($p=2.5{\times}10^{-7}$), while the \texttt{gpt-4o-mini} row is a genuine null rather than an untested one, and that model stays inside an $8$-point band across every property we varied (\S\ref{app:imgprops}). The comparison that matters is that colour, content and pixel count move refusal \emph{while an image is attached in both arms}, which presence-versus-absence contrasts cannot show. None of these properties carries information about the request, which is what makes the dependence a defect rather than a calibration.}
\label{tab:imgvsimg}
\end{table*}

\paragraph{What this still does not establish.}
Each arm remains a \emph{fixed} image held constant across prompts rather than sampled
per prompt, so we bound between-image variance along the five axes we varied and not
within them; natural photographs, complex scenes and adversarially chosen images remain
untested. One cell also decides whether the shift is universal across the hosted tier, and
it does not hold up: \texttt{gemini-2.5-flash} under the line drawing reads $+10$ points
($p=0.021$) in the smaller control and $+9$ points ($p=0.064$) in the larger property
sweep, the direction replicating across collection windows while the significance does
not. We therefore state the scope the stable cells support: \textbf{three of four hosted
models show the benign shift, and \texttt{gemini-2.5-flash} is near-null in both windows.}
The rates otherwise replicate: this job re-collected the blank-canvas arm from scratch and
reproduces rung~2 within a few points on every model (\S\ref{app:windows}), which matters
because these targets are \emph{not} deterministic at temperature $0$.

\paragraph{The instruction does not neutralise it.}
If the cue were under instructional control, a deployer could neutralise it for free. We test the cheapest possible mitigation: a neutral system message stating that the attached image is a fixed placeholder carrying no information about the request, and asking for it to be disregarded entirely (no safety framing, which could move refusal on its own).

\paragraph{What the instruction cue is made of.}
That inversion admits a deflationary reading, the model is reacting to the word ``image'' in its system prompt, and testing it requires a control the instruction arm does not provide: the arm it is compared against carries no system message \emph{at all}, so the contrast confounds the mention with the mere existence of a system prompt. We decompose it with a placebo ladder on \texttt{gemini-2.5-flash}: four arms in one job, each adding exactly one ingredient to the one below it, with \textbf{no image attached in any arm} (Table~\ref{tab:placebo}). A bare \texttt{Respond to the text of the request.} already costs $+10$ points ($p=0.013$).

\begin{table}[t]
\centering
\small
\begin{tabular}{llcc}
\toprule
 & system message & refusal & $\Delta$ \\
\midrule
A  & \emph{(none)} & $14\%$ &: \\
P2 & \texttt{Respond to the text\ldots} & $24\%$ & \textbf{+10}$^{*}$ \\
P1 & \quad+ ``any \emph{file} attached\ldots'' & $40\%$ & \textbf{+16}$^{**}$ \\
C  & \quad\emph{file} $\rightarrow$ \emph{image} & $39\%$ & $-1$ \\
\bottomrule
\end{tabular}
\caption{[\textsc{primary test, control arm}]~\textbf{Placebo ladder on \texttt{gemini-2.5-flash}}, $100$ prompts per arm, all four arms collected in one job, \textbf{no image attached in any arm}. Each rung adds one ingredient to the rung above, so the $+25$-point total ($p=4.2\times10^{-7}$) decomposes additively; P1 and C are byte-identical apart from the single word. $^{**}p<0.01$, $^{*}p<0.05$, exact McNemar paired per prompt. The C$-$P1 contrast is n.s.\ ($12$/$13$ discordant, $p=1.0$), which at $25$ discordant pairs excludes an image-word effect above $\sim\!11$ points: this model's response to an asserted attachment is not a response to the modality word.}
\label{tab:placebo}
\end{table}

\subsection{What it is not}
\label{sec:res-notwhat}

\paragraph{It is attachment, not the mention of an image.}
An image cannot be attached without the request context changing in other ways, so a natural objection is that the cue is lexical or contextual rather than visual: an objection our own \texttt{gemini-2.5-flash} result invites, since there a canvas alone does nothing while a canvas plus an instruction to disregard it adds $32$ points (Table~\ref{tab:placebo} resolves that cell; here we test the objection on the checkpoint where attachment does carry the effect). We test it directly with a $2\times2$ crossing attachment against a system-message mention of an image, worded so that the identical string is true whether or not an image is attached (Table~\ref{tab:factorial}).

\paragraph{It is not an input-length effect.}
A larger canvas costs more input tokens, so the size trend admits a reading with no visual content in it at all: the model may simply refuse more as its input grows, and pixel count may be standing in for token count. We test that by holding the token budget fixed and changing only the modality that fills it. \texttt{qwen3-vl-8b} encodes the $256^2$, $512^2$ and $1536^2$ canvases as $64$, $256$ and $2304$ visual tokens, so we build a text ladder matched to those budgets: the same canonical prompt padded with a topic-neutral filler sentence repeated $3$, $13$ and $110$ times, with no image attached and no attachment asserted.

\paragraph{Same weights, two serving stacks.}
The hosted-versus-open comparison varies six things at once (family, pretraining,
post-training alignment, vision encoder, inference implementation and vendor-side
filtering), so by itself it cannot say which carries the effect. One checkpoint cuts the
bundle in half. \texttt{gemma-3-12b-it} is open, vision-capable, and served both by a
managed commercial host and by our own vLLM, so the identical manipulation down both
routes holds weights, encoder, pretraining and alignment fixed \emph{by construction}. Managed minus self-served is $+1$, $+1$, $+1$ and $-3$ points, on one to three flips of
$100$ paired prompts, while the presence effect itself reproduces at $+7$ points on
\emph{both} routes.

\paragraph{It is not a property of vision--language models as such.}
Everything above is measured on commercial APIs. We repeated the identical manipulation (same canonical text, same blank canvas, no defense, paired per prompt id) on three open-weight VLMs served locally under vLLM with no moderation layer (Table~\ref{tab:ow_threshold}). The filler is constrained to contain no modality or attachment word, since padding that mentioned a file would reproduce the asserted-attachment cue of Table~\ref{tab:placebo} instead of isolating length. Measured on the target's own token accounting, the padded arms add $63$, $273$ and $2310$ tokens against the canvas arms' $66$, $258$ and $2306$, so at the decisive rung the text arm carries \emph{more} context than the image arm it is matched against. All seven arms run in one job over the same $100$ prompt ids, on request texts verified byte-identical across arms. The two ladders do not behave alike (Table~\ref{tab:tokenbudget}). The padded arms refuse at $47$, $50$ and $54\%$ against a $54\%$ text baseline: not one sits above the baseline at any budget, and the largest lands exactly on it ($8$ flips each way, $p=1.00$). The canvas arms at the same budgets refuse at $73$, $76$ and $86\%$. At the matched extreme the two differ by $32$ points on the same prompts, $33$ flips to $1$ ($p=4.1\times10^{-9}$), with the text side holding the larger budget. Input length is not what this model is charging for. One feature of the text ladder deserves stating rather than smoothing. It does rise, $47\!\to\!50\!\to\!54$, but it rises \emph{to} the baseline from below rather than above it, and the trend is not significant ($10$ flips to $3$, $p=0.09$). Our reading is that a short block of irrelevant context slightly relaxes refusal and a longer one returns it to baseline. What the data exclude is a positive length effect of the size the canvas produces: to explain the $1536^2$ arm by token count alone, $2310$ tokens of text would have to buy $32$ points, and they buy none.

\begin{table}[t]
\centering
\scriptsize
\setlength{\tabcolsep}{3pt}
\begin{tabular}{lccc}
\toprule
 & \multicolumn{2}{c}{benign refusal \%} & harmful \\
\cmidrule(lr){2-3}
Model & neutral & borderline & ASR \% \\
\midrule
qwen2.5-vl-7b & 0\,$\to$\,2 & 38\,$\to$\,37 & 4\,$\to$\,7 \\
internvl3-8b  & 1\,$\to$\,3 & 48\,$\to$\,55 & 5\,$\to$\,2 \\
pixtral-12b   & 0\,$\to$\,1 & \phantom{0}4\,$\to$\,\phantom{0}2 & 48\,$\to$\,\textbf{81}$^{***}$ \\
\bottomrule
\end{tabular}
\caption{[\textsc{exploratory follow-up}]~\textbf{Open-weight models, same manipulation} (text\,$\to$\,blank canvas), no defense, no moderation layer, $100$ prompts per cell, paired per prompt id, judged by \texttt{gpt-5-mini} as elsewhere. $^{***}p=1.0\times10^{-7}$, exact McNemar; all unmarked contrasts n.s. The benign inflation that reaches $+23$ to $+51$ points on hosted models (Table~\ref{tab:ladder}) is absent here, while \texttt{pixtral-12b} inverts on the harmful axis (\S\ref{sec:res-threshold}).}
\label{tab:ow_threshold}
\end{table}

\paragraph{But it does not require a hosted serving stack either, and a sample of open-weight nulls is what makes that easy to miss.}
The three nulls above are the whole open-weight evidence a three-model sample provides, and taken alone they license a conclusion the wider scan refutes: that the benign tax belongs to an aligned, \emph{moderated serving stack} rather than to the checkpoint. Scanning five open-weight VLMs under the identical protocol (Table~\ref{tab:tierscan}, Supplementary Document) turns up \texttt{qwen3-vl-8b} at $\mathbf{+32}$\textbf{ points} ($32$ discordant pairs to $0$, $p<10^{-4}$): an open, downloadable, $8$B checkpoint that we serve ourselves with nothing wrapped around it, showing an effect the size of the frontier hosted models. The nulls are therefore a property of which three models a small sample happens to contain, not of open weights.

\paragraph{The open checkpoint's cost is measured three times.}
Because this one checkpoint now carries the paper's scope claim, we do not rest it on a single collection. Its borderline-rung shift reads $+32$, $+28$ and $+29$ points across three independently collected jobs (Tables~\ref{tab:tierscan} of the Supplementary Document,~\ref{tab:generational},~\ref{tab:owladder}), against a text baseline that itself moves $51$/$54$/$53\%$ between them: the nondeterminism band that makes within-job pairing mandatory.

\paragraph{What the open checkpoint excludes, and what we leave open.}
Attaching an image changes four things at once: provider preprocessing, the serialisation of the request, whether the vision encoder runs, and how many visual tokens enter the context. On a hosted endpoint these cannot be taken apart. On \texttt{qwen3-vl-8b}, which we serve ourselves, three can be set aside.

\subsection{The price, and its decoupling}
\label{sec:res-price}

\paragraph{The benefit side, on a matched population.}
A A threshold shift should also catch more borderline-harmful requests. We measure that on OR-Bench's \emph{harmful} split: the constructed counterpart of the borderline-benign rung, same pipeline and topic distribution, so the two sides are the apples-to-apples comparison a differently-built harmful corpus would not give.

\paragraph{What the two sides do and do not license.}
Reporting a cost and a benefit beside each other invites a net-welfare reading that our design does not identify, so we state the arithmetic such a reading would require and leave its inputs to the deployer. Attaching the canvas pays for itself only if one additional harmful completion is worse than one additional false refusal by some factor $k$; at harmful-request prevalence $\pi$, the break-even factor is $k^{\star}=\frac{1-\pi}{\pi}\cdot\frac{\Delta_{\text{benign}}}{\Delta_{\text{harmful}}}$. The second term is a property of the model, and we measure it: $2.8$ to $6.8$ across the three hosted models with both sides.

\paragraph{The two sides do not move together.}
Both sides are measured on the two splits of \emph{one} benchmark built by one pipeline, that is the design, not an oversight: a differently-constructed harmful corpus would confound topic shift with the manipulation, and the matched pair is what makes the comparison interpretable at all. Placed side by side on the borderline rung, image presence costs $+51$, $+34$ and $+23$ points of benign refusal on \texttt{claude-sonnet-4-6}, \texttt{gpt-4o-mini} and \texttt{gemini-2.5-flash-lite}, and buys $18$, $5$ and $6$ points of prevented attack success on the matched harmful split (Table~\ref{tab:hostedharm}). \textbf{The benefit side collapses across the set while the cost side does not track it down}: prevented harm falls from $18$ points to $5$, while the benign price stays between $23$ and $51$. \texttt{gemini-2.5-flash} neither pays nor buys ($-3$ points and $-1$ points, both n.s.), and \texttt{qwen3-vl-8b} (\S\ref{sec:res-threshold}) pays $+29$ points against a $2\%$ harmful baseline that leaves almost nothing to prevent.

\begin{table}[t]
\centering
\scriptsize
\setlength{\tabcolsep}{3pt}
\begin{tabular}{lccrrr}
\toprule
Model & no image & $+$canvas & $\Delta$ & gain/loss & $p$ \\
\midrule
claude-sonnet-4-6     & 18 & \phantom{0}0 & $\mathbf{-18}$ & 0/18 & $\mathbf{7.6{\times}10^{-6}}$ \\
gemini-2.5-flash-lite & 11 & \phantom{0}5 & $\mathbf{-6}$  & 0/\phantom{0}6 & $\mathbf{0.031}$ \\
gpt-4o-mini           & \phantom{0}6 & \phantom{0}1 & $-5$ & 0/\phantom{0}5 & $0.063$ \\
gemini-2.5-flash      & 14 & 13 & $-1$ & 5/\phantom{0}6 & $1.00$ \\
\bottomrule
\end{tabular}
\caption{[\textsc{primary test}]~\textbf{What the canvas buys on the hosted models}: attack success (\%) on OR-Bench's harmful split, the constructed counterpart of the borderline-benign rung, under \{no image, blank canvas\}, no defense, $100$ prompts, paired per prompt id, HarmBench classifier on \texttt{gpt-5-mini}. ``gain/loss'' counts the discordant pairs in each direction on the same convention as Table~\ref{tab:headroom} of the Supplementary Document: a completion the canvas \emph{caused}, then one it prevented. \textbf{This is the buy side of the price reported beside it}, and it is small and getting smaller: the largest reduction is $18$ points on the model with the most headroom, two of the four do not clear $p<0.05$, and on \texttt{gemini-2.5-flash} the difference is a single prompt. Set against the $+51$, $+34$ and $+23$ points of benign cost on the same models (Table~\ref{tab:ladder}), and read together with the open checkpoint that pays $+29$ points against a $2\%$ harmful baseline (Table~\ref{tab:owladder} of the Supplementary Document), this is the decoupling the section reports. No prompt is shared between this table and the benign rungs; the two splits are matched by construction, not by item.}
\label{tab:hostedharm}
\end{table}

\emph{We deliberately do not reduce this to a single exchange rate.} Such a ratio is dominated by its denominator, and the denominator here is small and shrinking: $18$, $5$ and $6$ prevented completions out of $100$ prompts, in the same model order as above. A statistic that loses precision, and eventually becomes undefined, exactly where the trend it would summarise is strongest is the wrong instrument for that trend; on \texttt{gemini-2.5-flash} the harmful-side difference is already a single prompt, and on \texttt{qwen3-vl-8b} there is no rate to compute at all. Two caveats belong with the comparison rather than beneath it. It is conditional on an implicit $1{:}1$ harmful-to-benign prevalence; real traffic is overwhelmingly benign, so a deployment-relevant cost would scale the benign side up by the true prevalence ratio and make the trade substantially worse than anything above. And the comparison is not by itself a verdict: converting it into one requires a stated willingness to trade benign refusals against harmful completions, a deployment decision we do not make on the reader's behalf. We report both sides; we deliberately do not report a summed error count, which would silently weight the two error types equally.

\paragraph{On the open checkpoint the benign cost is paid against a floored harmful side.}
Running the full ladder plus the matched harmful set on \texttt{qwen3-vl-8b} in one job (Table~\ref{tab:owladder}) reproduces the paper's structural claim on a downloadable model: the neutral rung moves $+3$ points (n.s.) against the borderline rung's $+29$ points ($30$ discordant to $1$). The harmful side does something the hosted models could not show us. This checkpoint yields a harmful completion on only $2\%$ of plain harmful requests, below every hosted model in our set ($18$, $14$, $11$, $6\%$), so the canvas has essentially nothing left to prevent.

\paragraph{Given headroom, the cue does buy something, but not in a fixed direction.}
We rebuilt the harmful side on the same checkpoint, lifting the floor by attacking rather than by switching models, so that serving stack, canvas, judge and target are all held fixed and only the amount of available harm changes. On $100$ HarmBench behaviours, four encodings place the no-image arm at $17$, $52$, $51$ and $63\%$ attack success, clear of the $2\%$ floor; each arm was then re-run with the identical $512^2$ white canvas appended and byte-identical encoded text on both sides, so the attachment is again the only difference (Table~\ref{tab:headroom}). \textbf{The result withdraws the stronger reading: the cue is not inert once there is headroom.} It lowers attack success on three of the four arms, significantly on two ($-10$ points, $p=2.0\times10^{-3}$; $-12$ points, $p=2.9\times10^{-2}$), and it \emph{raises} it on the fourth ($+10$ points, $p=5.2\times10^{-2}$). What the four arms share is the request; what differs is how it is written.

\paragraph{The decoupling holds inside one family; the \emph{trend} cannot be tested there.}
A reader is entitled to ask whether the asymmetry is an artifact of comparing unlike
models, since family, alignment procedure, base refusal rate and judge all vary across
our set at once. We therefore measured the matched harmful split on all three rungs of
the generational ladder, where every one of those is held fixed and only the checkpoint
moves (Table~\ref{tab:generational} of the Supplementary Document). The decoupling survives the control: at each rung
the harmful effect is a bounded null, with $95\%$ intervals $[-4.5,+4.5]$, $[-2.7,+9.3]$
and $[-6.1,+3.7]$ points, all inside a $\pm10$-point equivalence margin, while the benign cost
runs $+8$, $+1$ and $+28$ points.

\paragraph{The behaviour arrives as a step, not a gradient.}
An open family lets us ask \emph{when} the behaviour appeared. Measured in one job against one canonical text, three generations of the same family give $+8$ points (n.s.), $+1$ points, and $+28$ points ($p<10^{-4}$) from oldest to newest (Table~\ref{tab:generational}). That multimodal safety behaviour is this movable is consistent with work on the training side: \citet{gulati2026narrowfinetuning} find that narrow downstream fine-tuning of an aligned VLM shifts safety behaviour sharply and non-locally, and, relevant to why this paper measures the image channel separately at all, that multimodal evaluation reveals substantially more degradation than text-only evaluation of the same checkpoint.

\paragraph{The sign is not fixed either.}
The harmful axis then does something we did not predict. On \texttt{pixtral-12b} the
blank canvas moves ASR $48\!\to\!81\%$ ($+33$ points, $95\%$ CI $[+21.8,+43.0]$, $37$
prompts flipping toward harm against $4$ away, $p=1.0\times10^{-7}$): the same
request-independent image, on the same axis, moving safety hard in the \emph{opposite}
direction. We then tested a checkpoint chosen because \citet{wei2026representationshift} already predicted the direction, reporting $+28$ points from a blank image on \texttt{llava-1.5-7b}. \S\ref{app:outcome} notes where our binary refuse/comply outcome is coarser
than the response taxonomies of \citet{ren2026seeingthreat} and
\citet{yang2026refusalreframing}.

\paragraph{Taken together.}
The surfaces make the cue look worse, not better, as a safety signal. Image presence tightens the threshold on three hosted models, loosens it by $33$ and $39$ points on two open-weight models, and does nothing on a fourth hosted model or on two other open-weight ones: under one manipulation that carries zero task information.

\section{Discussion}
\label{sec:discussion}

\paragraph{What we can and cannot say about the mechanism.}
Our central result is behavioural: a request-independent image moves the refusal threshold, it does so for two entirely different request-independent image classes, it varies with properties of the attached file that carry no information about the request, and it survives an explicit instruction to disregard the image. A black-box threat model gives us no way to inspect activations or vendor pipelines, so we can characterise the effect precisely without confirming a mechanism from the inside. Two interpretability findings bear on why a channel carrying nothing could matter, cited as hypotheses since neither studies safety: models treat a textual concept with no visual evidence as though the image contained it~\citep{kim2025visualabsence}, and visual information is retrieved weakly and suppressed later, so answers survive severe obfuscation~\citep{zhou2026visualignorance}.

\paragraph{Where this sits among multimodal safety findings.}
Work on images and VLM safety divides into five kinds, and our contribution is legible
only against the split. \emph{(i) Intrinsic modality shifts} ask what the visual channel
does to a model's own safety behaviour~\citep{zou2026understanding,gupta2026safetyoverrides}. \emph{(ii) Visual jailbreaks} make the image carry or compose the attack, as rendered text, cross-channel composition, optimised images, scene construction, symbol ciphers, or universal transferable patterns~\citep{gong2025figstep,chen2026textdj,shayegani2023jailbreakpieces,kim2025benigntotoxic,miao2025visualcontextual,azulay2026visualmodality,cui2026ultrabreak}. \emph{(iii) Safety fine-tuning and test-time guardrails} add a protective layer, through safety-preference training, borderline-case diagnosis, adversarially-aware preference optimisation, inference-time visual-safety mechanisms, persona induction, and visual-token risk amplification~\citep{zong2024vlguard,zhang2025spavl,lee2026holisafe,ding2026rethinkingbottlenecks,weng2025adversaryawaredpo,zhang2026davsp,yang2026visualselffulfilling,wang2026riskawareness}. \emph{(iv) Over-refusal measurement} asks what alignment costs on traffic that was never harmful, over dual-use imagery, in-the-wild images, image-seeded multi-turn conversations, high-stakes text domains, and unified safety-utility benchmarks~\citep{ren2025dualbench,lee2025meme,jindal2025reveal,zhang2026healthorsc,palaskar2026vlsu,geng2025vscbench}. \emph{(v) Black-box
routing artifacts} ask what the deployment path itself does to a safety
decision~\citep{plebe2026imagesamplify}. The over-refusal line is closest in
spirit, but it measures the cost of alignment over image \emph{content}; ours is the cost
of the attachment interface itself, and the benchmarks in that
line~\citep{palaskar2026vlsu,geng2025vscbench} score a model's safety-utility trade
without separating the presence axis out of it.

\paragraph{Shortcut-driven safety judgements.}
A parallel line reaches a conclusion close to ours from a different manipulation. \citet{hinojosa2026saves} study embodied situational safety and show judgements track the \emph{semiotics of a marker} rather than the hazard: a red circle drives refusal to $73.1\%$ where a white one on the same region gives $41.8\%$. Their conclusion, that safety decisions
rest on learned visual--linguistic associations rather than grounded visual
understanding, converges with ours, and it is worth being precise about where
the two differ, because the difference is our contribution.

\paragraph{Relation to safety-perception distortion.}
The closest prior work is \citet{zou2026understanding}, who use the same blank-image control and argue on that basis that the shift originates in the visual modality rather than in image content; our design choice is theirs, and our two-image control extends the same argument. Their analysis is directed at the direction in which distortion \emph{weakens} safety, the opposite of the tightening we measure on hosted models, and our open-weight arm resolves that as a genuine sign difference rather than a disagreement: the same manipulation tightens on four moderated hosted models and loosens $33$ points on \texttt{pixtral-12b}, while \texttt{pixtral-12b} and \texttt{qwen3-vl-8b} are both open checkpoints under one serving arrangement and move in opposite directions.

\section{Conclusion}
\label{sec:conclusion}

Safety-aligned vision--language models shift their refusal threshold on whether an image
is attached, a property of the request's \emph{form} carrying no per-prompt information
about what is asked. Measured with no defense anywhere in the pipeline, the shift is large
($+23$ to $+51$ points on the borderline rung), it is a threshold shift rather than
blanket caution, it survives an explicit instruction to disregard the image, it survives
holding the mention of an image byte-identical across arms, and on one checkpoint it fires
on an asserted attachment with nothing attached.

\section*{Limitations}

Table~\ref{tab:claims} of the Supplementary Document sorts every claim in this paper by
what supports it: a direct paired measurement, a control with a stated bound, or neither.
\S\ref{app:limits} elaborates the entries below.

\paragraph{Scope.} We do not claim the effect is universal. We describe four hosted models and seven
open-weight checkpoints and report the models on which it is absent as prominently as
those on which it is large; three of the seven show no benign cost at all, and a sample
of that kind invites a claim about model classes that a fourth open checkpoint refutes. The positive claim is therefore narrow: the behaviour is a property of \emph{particular
aligned checkpoints}, hosted or open, and no label we have tried predicts which.

\paragraph{Coverage.} The full ladder plus matched harmful set is complete on four hosted models and on the one
open checkpoint carrying the scope claim; the other six open checkpoints are measured on
the borderline rung only. The generational ladder is one family at one size class, and
because it is already floored on the harmful split at its oldest rung ($2\%$) it can
demonstrate the decoupling under control but cannot test the \emph{trend} in harmful
headroom, which stays an across-model observation. The route control is run on a
small-effect checkpoint, so its bound sits at roughly the scale of the effect it
validates; the stronger design is not currently available to us
(\S\ref{app:routecontrol}).

\paragraph{Instrument.} Refusal and harm are assigned by LLM judges. Our claims are paired within-prompt
contrasts scored by a held-constant instrument, which is the property they depend on, and
three judges have now been applied to these cells with every effect keeping its sign,
significance and approximate magnitude (\S\ref{sec:res-judge}). What remains untested is a
blind spot common to all rubric-following language models, which only the two human
anchors speak to and which they bound only at $n\approx50$ per arm.

\section*{Ethics Statement}

All experiments use standardized public benchmarks (HarmBench, JailbreakBench, OR-Bench, AlpacaEval) released for safety evaluation. \textbf{We introduce no attack.} The only manipulation is attaching a request-independent image to an otherwise unchanged request, and nothing here increases what an attacker can do that attaching an arbitrary image did not already permit. We judge disclosure net-positive, since the cost falls on benign users who cannot avoid it without being told it exists (\S\ref{app:ethics}).

\bibliography{paper}

\clearpage
\appendix
\setcounter{table}{0}
\setcounter{figure}{0}
\renewcommand{\thesection}{S\arabic{section}}
\renewcommand{\thefigure}{S\arabic{figure}}
\renewcommand{\thetable}{S\arabic{table}}

\begin{center}
{\LARGE\bfseries Supplementary Material\par}
\end{center}
\vspace{1em}

\noindent This supplementary material supplements the main paper. It carries sample
stimuli, pre-specified predictions and ablation counts, statistical detail,
the multiplicity accounting, the wider image-property sweep and the
reproducibility record that a reader would consult to \emph{check} a result
reported in the main paper.

\vspace{0.5em}
\noindent The main paper states in its own text every quantity its argument turns
on. What is collected here is the evidence a reader consults to \emph{check} those
quantities, together with the protocol, the corrections, and the supporting arms
that qualify or replicate a claim rather than establish it. Eight tables cited by
the main paper are printed here: the alignment-tier scan
(Table~\ref{tab:tierscan}), the generational ladder (Table~\ref{tab:generational}),
the token-budget control (Table~\ref{tab:tokenbudget}), the attachment-by-mention
factorial (Table~\ref{tab:factorial}), the topic-stratified cost
(Table~\ref{tab:strata}), the open-checkpoint ladder (Table~\ref{tab:owladder}),
the harmful-side headroom table (Table~\ref{tab:headroom}), and the index sorting
every claim by what supports it (Table~\ref{tab:claims}).

\vspace{0.5em}
\noindent Section, figure and table numbers here are prefixed \textbf{S}.
References of the form ``Table~1'' point into the main paper.

\vspace{1em}

\section{Sample images and variant illustration}
\label{app:sample_images}

\begin{figure*}[t]
\centering
\includegraphics[width=\textwidth]{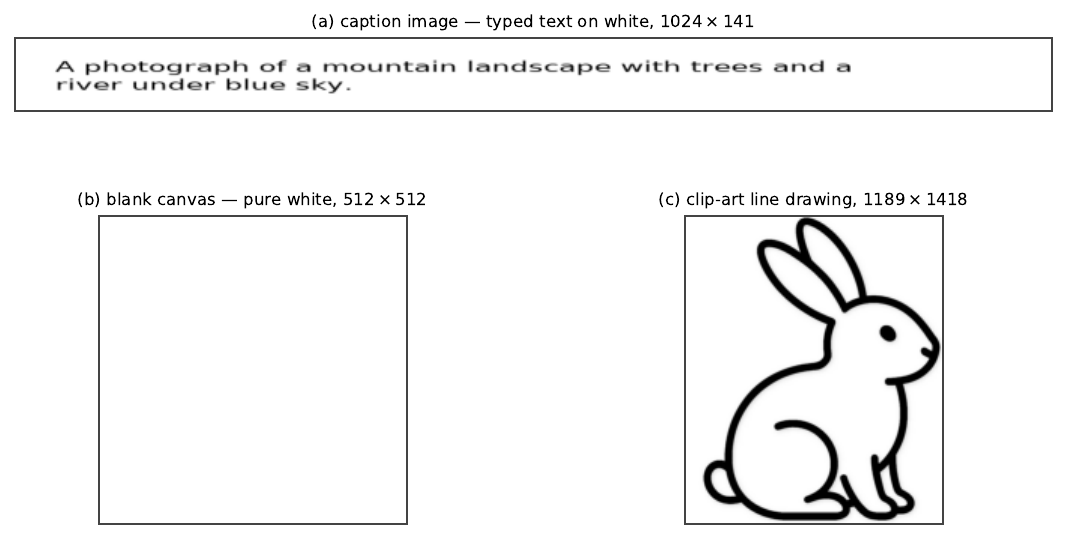}
\caption{The three request-independent images used in this paper. \textbf{(a)} the caption image: a $1024\times141$ PNG of typed text on white, the only arm carrying readable characters, which is what places an OCR path on one side of the comparison. \textbf{(b)} a $512\times512$ pure-white blank. \textbf{(c)} a $1189\times1418$ clip-art line drawing \emph{of a rabbit}: drawn content, no readable text, and no relation to any request in either benchmark. We name the subject because ``line drawing'' alone leaves a reader unable to check that the drawn content is request-independent, which is the property the arm depends on. None relates to any request, and within an arm every prompt receives the byte-identical file (\S\ref{sec:method}).}
\label{fig:decoys}
\end{figure*}

\paragraph{Caption image.} The OCR-able request-independent image used in the property contrasts is a $1024\times141$ PNG containing the typed text ``A photograph of a mountain landscape with trees and a river under blue sky.'' on white (Fig.~\ref{fig:decoys}(a)): a caption rendered as an image, not a photograph. It is the only request-independent image we test that carries readable text, which is what lets us place an OCR path on both sides of the comparison. Fig.~\ref{fig:decoys}(b) is the $512\times512$ pure-white blank and (c) the clip-art line drawing.

\section{Predictions, ablation counts, and judge robustness}
\label{app:counts}

\paragraph{Per-cell counts.}
Every cell in this paper is the average over $100$ paired prompts (rows $0$--$99$). All arms in a contrast are chained off one canonical text step, so the text channel is byte-identical across them (\S\ref{sec:method}, Pairing protocol) and within-cell variance comes from the target model's own sampling alone. Decoding is greedy throughout (temperature $=0$).

\paragraph{Inference parameters.}
All LLM calls use greedy decoding: temperature $0.0$ with \texttt{max\_tokens}\,$=16{,}384$. Remaining sampling parameters (top-$p$, top-$k$, frequency/presence penalties) are left at provider defaults, which are no-ops under greedy decoding. No random seed is pinned, two of the three providers used here do not expose one, so repeated runs are reproducible in distribution rather than bit-for-bit; every number we report comes from a single pinned run. This applies to the target VLM and to the judge alike.

\paragraph{Provenance of every headline result.}
Table~\ref{tab:provenance} states, for each result in this paper, its collection window, target models and judge backbone. Three things can be read off it directly. First, \textbf{no cell in this paper has a defense in the loop}: the evaluation-protocol distinctions that would otherwise need stating do not arise here. Second, every ASR and refusal number is scored by \texttt{gpt-5-mini}; no cell here was collected under a different backbone and left unrescored. Third, rows collected on different dates are never compared against one another: every contrast we report lives inside a single job, for the nondeterminism reason given in \S\ref{sec:method}.

\begin{table*}[t]
\centering
\scriptsize
\setlength{\tabcolsep}{4pt}
\begin{tabular}{lllll}
\toprule
Result & Collected & Target models & Judge(s) & Cells \\
\midrule
\multicolumn{5}{l}{\emph{Hosted models}}\\
Tab.~\ref{tab:ladder} (benign sensitivity ladder) & 2026-08-06/08 & 4 hosted & \texttt{gpt-5-mini}, \texttt{gpt-5-nano} & 32 \\
Tab.~\ref{tab:hostedharm} (hosted harmful side) & 2026-08-06/08 & 4 hosted & \texttt{gpt-5-mini}, \texttt{gpt-5-nano} & 16 \\
Tab.~\ref{tab:strata} (category-stratified cost) & 2026-08-09 & claude, fl.-lite, 4o-mini & \texttt{gemini-2.5-pro}, \texttt{gpt-5-mini} & 12 \\
Tab.~\ref{tab:presence} (presence controls) & 2026-08-07/08 & 4 hosted & \texttt{gpt-5-mini}, \texttt{gpt-5-nano} & 32 \\
Instruction arm, \texttt{gemini-2.5-flash} & 2026-08-08 & flash & \texttt{gpt-5-mini} & 4 \\
Tab.~\ref{tab:placebo} (asserted-attachment ladder) & 2026-08-08 & flash & \texttt{gpt-5-mini} & 4 \\
Tabs.~\ref{tab:imgvsimg},~\ref{tab:imgprops},~\ref{tab:contentinstance} (image properties) & 2026-08-07/08 & 4 hosted & \texttt{gpt-5-mini}, \texttt{gpt-5-nano} & 74 \\
\addlinespace[2pt]
\multicolumn{5}{l}{\emph{Open-weight models, self-served under vLLM with no moderation layer}}\\
Tab.~\ref{tab:sameweights} (serving-route control) & 2026-08-07 & \texttt{gemma-3-12b-it}\,$\times$\,2 routes & \texttt{gpt-5-mini} & 8 \\
Tab.~\ref{tab:ow_threshold} (first open-weight arm) & 2026-08-06/09 & 3 open-weight & \texttt{gemini-2.5-pro}, \texttt{gpt-5-mini}, \texttt{gpt-5-nano} & 40 \\
Tab.~\ref{tab:tierscan} (five-model scan) & 2026-08-08 & 5 open-weight & \texttt{gpt-5-mini} & 10 \\
Tab.~\ref{tab:owladder} (ladder + harmful, open ckpt) & 2026-08-08/09 & \texttt{qwen3-vl-8b} & \texttt{gemini-2.5-pro}, \texttt{gpt-5-mini} & 8 \\
Tab.~\ref{tab:instance} (size\,$\times$\,fill replication) & 2026-08-08 & \texttt{qwen3-vl-8b} & \texttt{gpt-5-mini} & 11 \\
Tab.~\ref{tab:factorial} (attachment\,$\times$\,mention) & 2026-08-08 & \texttt{qwen3-vl-8b} & \texttt{gpt-5-mini} & 4 \\
Tab.~\ref{tab:generational} (generational ladder, benign) & 2026-08-08 & 3 open-weight & \texttt{gpt-5-mini} & 6 \\
Tab.~\ref{tab:generational} (generational ladder, harmful) & 2026-08-09 & \texttt{qwen2-vl-7b} & \texttt{gpt-5-mini} & 2 \\
Tab.~\ref{tab:headroom} (harmful side, floor removed) & 2026-08-21 & \texttt{qwen3-vl-8b} & \texttt{gpt-5-mini} & 8 \\
Tab.~\ref{tab:tokenbudget} (padded-text control) & 2026-08-21 & \texttt{qwen3-vl-8b} & \texttt{gpt-5-mini} & 7 \\
\texttt{llava-1.5-7b} sign inversion (\S\ref{sec:discussion}) & 2026-08-09 & \texttt{llava-1.5-7b} & \texttt{gpt-5-mini} & 4 \\
\bottomrule
\end{tabular}
\caption{\textbf{Provenance of every result in this paper.} One row per collection
campaign; the ``Cells'' column counts the released per-prompt judgment files behind
it, and the eighteen rows sum to the $282$ cells shipped in the code artifact's
judgments index, which is the machine-readable form of this table and carries the
exact run directory, timestamp, prompt count and judge for each cell individually.
\textbf{No cell in this paper has a defense in the loop}, so the defense-configuration
and evaluation-protocol columns that would ordinarily appear here are vacuous and are
omitted: the only manipulation anywhere is whether an image is attached. Benign rungs
are scored with the JailbreakBench refusal classifier and harmful arms with the
HarmBench classifier. A row naming more than one judge ships the original
\texttt{gpt-5-mini} scoring \emph{plus} the robustness rescores of
\S\ref{sec:res-judge}, which re-score stored responses and re-query no model. Every
contrast is collected inside a single job (\S\ref{sec:method}); rows collected on
different dates are never compared to one another, and a date range spans the arms of
one campaign rather than the two sides of one contrast. Three cells of the
randomised-interleaving control were stored without per-prompt rows and ship as
recorded rates only; they are the sole exception and are flagged in the index.}
\label{tab:provenance}
\end{table*}

\paragraph{Judge standardization and per-row audit trail.}
ASR and refusal judges are the standardized HarmBench classifier and JailbreakBench-refusal classifier respectively, both peer-reviewed evaluators distributed with their respective benchmarks, with empty-response auto-classification (\S\ref{sec:setup}) handling upstream API blanks correctly. Every number in this paper is scored on the \texttt{gpt-5-mini} backbone. Where a result predates that choice, all stored responses were re-scored with \texttt{gpt-5-mini} without re-querying any target; no claim in this paper depends on that re-scoring. Every per-prompt evaluation row stores the full judge audit trail (\texttt{judge\_output}, \texttt{judge\_reasoning} and \texttt{judge\_raw\_response}) alongside the model response in \texttt{raw\_results.jsonl}, so any individual judgement can be inspected without re-running the judge.

\section{Statistical tests}
\label{app:stats}

Because all arms in a contrast are generated from one canonical text (\S\ref{sec:method}), outcomes are paired at the prompt level. We therefore test every contrast with an \emph{exact two-sided McNemar test} on the discordant pairs, and report Wilson score $95\%$ intervals for every reported proportion (Wilson intervals are preferred to normal-approximation intervals here because several cells sit near $0$ or $100\%$, where the normal approximation misbehaves). Intervals on a \emph{difference} of paired proportions are computed by Newcombe's method~10, the paired-design counterpart of the Wilson interval, and every $\Delta$ interval quoted in the main paper is of that kind; the two are named separately throughout because they are intervals on different quantities. Exact $p$-values and discordant-pair counts are reported inline in every results table, rather than collected separately. Because an exact McNemar $p$ is a function of the discordant counts alone, printing $(b,c)$ makes every test in this paper independently recomputable from the published numbers. Multiplicity is treated separately in \S\ref{app:multiplicity}.

\section{Multiplicity}
\label{app:multiplicity}

This paper reports many significance tests, so we state in advance which family
each belongs to and report what a correction does to every claim. Families are
defined by \emph{claim}, not by table: one family per thing the paper asserts
(Table~\ref{tab:families}). Only \textbf{F1} is confirmatory; the rest are exploratory and are labelled as
such in the text as well as here. We apply Holm--Bonferroni (family-wise error)
within each family, except for the ten-arm property sweep, where the question is
which arms move at all rather than whether any single one does, and
Benjamini--Hochberg (false discovery rate) is the appropriate control.

\paragraph{What was fixed in advance, and what was not.}
\textbf{No multiplicity family in this paper was fixed in advance}: the families of Table~\ref{tab:families} were defined after the measurements existed. Individual read-outs were pre-specified (recorded in the repository before the job that measured them ran, which is what we mean by the term everywhere in this paper, and not registration with any external registry) and one such prediction failed and is reported as failed above; what was never pre-specified is the family structure the corrections below are computed over. We state
that plainly because it changes how two results should be read. First,
\texttt{qwen3-vl-8b} was \emph{selected}: it is the one checkpoint of five in
the open-weight scan (Table~\ref{tab:tierscan}) that showed a large effect, so
the property sweep, generational ladder, attachment\,$\times$\,mention factorial
and harmful side that follow are all follow-up on a selected model rather than
independent confirmation of the scan. What protects them is not the scan but
re-collection: the presence effect on that checkpoint reproduces at $+32$, $+28$
and $+29$ points across three independently collected jobs, and a selection artifact
would not survive being measured again. Second, the property sweep was
exploratory by construction (we did not know which axis, if any, would carry
the price), which is why it is controlled for false discovery rate, and why we
treat the surviving axis (image size) as a finding to be replicated rather than
as an established property of images. \textbf{F1} is confirmatory only in the
weaker sense that its contrast was specified before the job that measured it,
not that it was registered anywhere.

\begin{table}[h]
\centering
\scriptsize
\begin{tabular}{llcc}
\toprule
 & family & $n$ & method \\
\midrule
F1 & presence effect, borderline rung (\textbf{primary}) & $10$ & Holm \\
F2 & which image properties change the price & $10$ & BH \\
F3 & generational ladder & $3$ & Holm \\
F4 & instruction/mention decomposition & $6$ & Holm \\
F5 & ladder rungs other than borderline & $3$ & Holm \\
F6 & size axis, replicated across three fills & $3$ & Holm \\
F7 & sign inversion on the harmful axis & $2$ & Holm \\
F8 & benign cost by OR-Bench category, stratified & $30$ & BH \\
F9 & direct between-arm property contrasts & $2$ & Holm \\
\bottomrule
\end{tabular}
\caption{\textbf{Declared test families.} F1 pools the borderline-rung presence
contrast across all ten models we measured it on, hosted and open, with no
selection: the four hosted models of Table~\ref{tab:ladder}, the five of
Table~\ref{tab:tierscan}, and \texttt{llava-1.5-7b}, including every null and
both signs. F7 exists because the sign inversion had previously been a lone
test belonging to no family, which left the paper's \emph{steerable} claim the
only one not carried through a correction.}
\label{tab:families}
\end{table}

\paragraph{What correction changes: two tests of sixty-nine.}
Across all nine families, exactly two results change status, both inside the primary family and both small.
\texttt{gemma-3-12b-it}'s $+9$ points on the borderline rung ($10/1$ discordant,
exact $p=0.0117$) and \texttt{llava-1.5-7b}'s $-8$ points ($1/9$, $p=0.0215$) are
significant uncorrected and do \emph{not} clear Holm inside the ten-test
primary family. We report both as non-significant accordingly. No claim in the
paper rests on either: the serving-route arm \texttt{gemma} appears in is an
\emph{equivalence} result about two hosting routes whose positive claim is carried
by \texttt{qwen3-vl-8b}, and \texttt{llava}'s load-bearing result is its harmful-axis
inversion ($p=3.6\times10^{-12}$), not its benign movement. Every other test keeps
its status: each significant result stays significant within its family, and
every null stays null.

\paragraph{The primary claim survives the most hostile correction available.}
Bonferroni across all $69$ corrected tests at once, ignoring the family
structure entirely and stricter than any reviewer would require, sets
the threshold at $\alpha/69=7.3\times10^{-4}$. The three hosted models with the
effect ($p\leq1.5\times10^{-6}$) and the open checkpoint
($p=4.7\times10^{-10}$) all clear it, as do the factorial's two attachment
contrasts, the generational step, and eight of the ten property arms.
Both harmful-axis sign inversions clear it as well, as do both direct property contrasts of F9 and $17$ of the $30$ stratified per-category tests. Thirty-six of the sixty-nine tests survive a global Bonferroni.

\paragraph{Correction does not rescue a null, and we do not ask it to.}
Three groups of results in this paper are negative, and a multiplicity
correction can only make a null more null. Their interpretation is governed by
power, not by $\alpha$, so we state the bound rather than the $p$ in each case:
the serving-route contrasts ($1$--$3$ discordant of $100$, $p\geq0.25$) support
equivalence between hosting routes; the image-word placebo contrast ($12/13$
discordant, $p=1.0$) excludes an effect above ${\sim}11$ points at $25$ discordant
pairs; and the three open-weight nulls are reported as facts about the models we
sampled rather than as a property of open-weight VLMs: a reading this paper
explicitly retracts (\S\ref{sec:res-threshold}), since a fourth open checkpoint
refuted it.

\paragraph{Reproducibility of the correction.}
The correction is computed by \texttt{src/analysis/paper\_b\_multiplicity.py} in
the released code, which holds the family declarations and the discordant counts
as read from the tables above. Because an exact McNemar $p$ depends only on
$(b,c)$, that script recomputes each test from the published counts rather than
from stored responses, and its audit mode cross-checks every recomputed value
against the $p$ typeset in this paper: all $19$ tests for which both are
available agree, with no mismatches.

\section{Reproducibility}
\label{app:repro}

\paragraph{Model identifiers.}
Hosted models are named by the identifier passed to the provider API:
\texttt{claude-sonnet-4-6} (Anthropic), \texttt{gpt-4o-mini} (OpenAI),
\texttt{gemini-2.5-flash} and \texttt{gemini-2.5-flash-lite} (Google). Judges run
on \texttt{gpt-5-mini} (OpenAI). Open-weight checkpoints are given by their
Hugging Face repository (Table~\ref{tab:modelids}), which is what a
replication should pull:

\begin{table}[h]
\centering
\scriptsize
\begin{tabular}{ll}
\toprule
paper name & Hugging Face repository \\
\midrule
\texttt{qwen2-vl-7b}          & \texttt{Qwen/Qwen2-VL-7B-Instruct} \\
\texttt{qwen2.5-vl-7b}        & \texttt{Qwen/Qwen2.5-VL-7B-Instruct} \\
\texttt{qwen3-vl-8b-instruct} & \texttt{Qwen/Qwen3-VL-8B-Instruct} \\
\texttt{internvl3-8b}         & \texttt{OpenGVLab/InternVL3-8B} \\
\texttt{pixtral-12b}          & \texttt{mistralai/Pixtral-12B-2409} \\
\texttt{llava-1.5-7b}         & \texttt{llava-hf/llava-1.5-7b-hf} \\
\texttt{gemma-3-12b-it}       & \texttt{google/gemma-3-12b-it} \\
\bottomrule
\end{tabular}
\caption{\textbf{Open-weight checkpoints.} The serving-route control pairs the
last row's self-served route against the same weights on AWS Bedrock
(\texttt{google.gemma-3-12b-it}); the two identifiers differ by one character,
so the pairing is carried in the model registry as a fact rather than inferred
from the strings.}
\label{tab:modelids}
\end{table}

\paragraph{Decoding and serving.}
Every target and judge call uses temperature $0$, \texttt{top\_p} $1.0$,
\texttt{top\_k} $0$, no frequency or presence penalty, a single completion, and a
$16{,}384$-token output cap. Our harness sets \texttt{seed}\,$=42$ where the
endpoint accepts one, but none of the three hosted providers exposes a
honoured seed, so the field is inert for every hosted number in this paper and
we do not claim bit-for-bit reproducibility anywhere (\S\ref{app:stats}). Open-weight models are served with vLLM
$0.25.0$ on PyTorch $2.11.0$ (CUDA $13.1$) and Transformers $5.13.1$, one server
per model, with each model's default chat template. We note again that
temperature $0$ does \emph{not} make the hosted targets deterministic
(\S\ref{sec:res-threshold}); this is why every contrast in this paper is
collected inside a single job and why cross-window comparisons are reported as
replications rather than as identities.

\paragraph{Parameter selection.}
No hyperparameter search was performed for this paper. The decoding settings
above were fixed \emph{a priori} at their greedy values and never tuned: the
selection criterion was reproducibility, not performance, since a tuned
decoding setting would make the paired contrast depend on a choice we had
optimised. One value was therefore tried per parameter. The sweeps this paper
does report vary experimental \emph{conditions} (image properties, rung, target
checkpoint), never model hyperparameters.

\paragraph{Image construction.}
Within an arm every prompt receives the byte-identical file, verified as exactly
one distinct image hash per condition before each run. Canvases are generated
programmatically from (width, height, fill), so every canvas arm is fully
specified by the parameters printed in its row and no image asset needs to
be distributed for those to be reproduced. The line drawing and caption images
are included in the supplementary archive.

\paragraph{System messages, verbatim.}
No arm carries a system message except where stated. The instruction-mitigation
arm (\S\ref{sec:res-threshold}) uses, with no safety framing that could move
refusal on its own:

\begin{quote}\small\ttfamily
An image is attached to the user's message. It is a fixed placeholder that
carries no information about the request. Disregard the image entirely and
respond to the text exactly as you would if no image were attached.
\end{quote}

\noindent The placebo ladder (Table~\ref{tab:placebo}) uses three messages, each
adding one ingredient. Rung P2: \texttt{Respond to the text of the request.}
Rung P1 prepends \texttt{Any file attached to the user's message is a fixed
placeholder that carries no information about the request.} Rung C is
byte-identical to P1 with the single word \emph{file} replaced by \emph{image}.
Rung A carries no system message at all, which is forwarded to the provider as
the absence of a system prompt rather than as an empty one: the distinction
that the ladder exists to control for.

\paragraph{Artifact statement.}
The Code and Data Supplement accompanying this submission contains the experiment
code, the YAML configuration for every run reported here, the analysis scripts
that produce each table and statistic (including the multiplicity correction of
\S\ref{app:multiplicity}), and the per-prompt evaluation rows for all $282$ cells
behind this paper, indexed by target model, defense, transformation chain, judge
model and rubric. Each row carries the per-prompt verdict together with the
judge's output, reasoning and raw response, so any individual judgment can be
inspected and any table recomputed without re-querying a target model. Every
released cell was checked to reproduce the metric recorded at collection time
before the archive was written, and the emitter refuses to release a cell that
does not.

Two limits on that archive, stated rather than left to be discovered. First, the
$46$ cells scored under the harm rubric hold working jailbreak completions to
public harmful behaviours; for those the completion text is withheld and its
length given instead, while the judge's verdict, reasoning and raw response all
ship, so the \emph{judgment} remains auditable and only the harmful text is not
redistributed. Those cells are flagged in the index. Every benign-rubric cell,
which is what each headline number in this paper is computed from, ships in full.
Second, three cells of the randomised-interleaving control (\S\ref{sec:method})
were stored without per-prompt rows and are therefore represented by their
recorded rates alone. Prompt sets are the public splits of OR-Bench,
JailbreakBench and HarmBench cited in \S\ref{sec:setup}; we redistribute the
exact slices used, identified by index, rather than the benchmarks themselves.

\section{The benign cost, stratified across topics}
\label{app:strata}

The borderline rung of the main paper draws its prompts from the head of OR-Bench's hard split, and that file is ordered by category, so those prompts are \emph{deception} and \emph{harassment} only. We therefore re-ran that rung on a category-stratified sample: $30$ prompts from each of the ten categories, $300$ in total, same manipulation, same rubric, both arms in one job per model. All thirty per-category contrasts move in the same direction and $26$ survive a Benjamini--Hochberg correction across the family. The effect is also \emph{larger} than the unrepresentative slice, so correcting the sample raises the measured cost rather than lowering it. Which topic pays most is strongly model-specific and shares no ordering across models.

\section{Control arms: the asserted attachment, and model selection}
\label{app:controls}

Two control arms referenced from the main paper. The first decomposes the one model on which a neutral instruction to disregard the image \emph{increases} refusal. Each rung adds one ingredient to the rung above, and \textbf{no image is attached in any arm}: what that model responds to is being told an attachment exists, not the attachment and not the word for it (replacing ``image'' with ``file'' is worth $-1$ points, n.s., which at $25$ discordant pairs excludes an image-word effect above $\sim\!11$ points). The second records how the open checkpoint carrying this paper's scope claim was found, and that a coarse alignment-tier label does not predict the effect: four of five models share one label and span $+32$ to $+0$ points.

\begin{table}[t]
\centering
\scriptsize
\setlength{\tabcolsep}{3pt}
\begin{tabular}{lccccc}
\toprule
Model & text & blank & $\Delta$ & disc. & $p$ \\
\midrule
\textbf{qwen3-vl-8b} & 51 & \textbf{83} & \textbf{+32} & 32/0 & $\mathbf{<10^{-4}}$ \\
gemma-3-12b-it                & \phantom{0}4 & 13 & +9 & 10/1 & 0.012 \\
internvl3-8b                  & 48 & 55 & +7 & 13/6 & 0.167 \\
qwen2.5-vl-7b                 & 37 & 37 & +0 & \phantom{0}7/7 & 1.000 \\
pixtral-12b                   & \phantom{0}4 & \phantom{0}3 & $-1$ & \phantom{0}3/4 & 1.000 \\
\bottomrule
\end{tabular}
\caption{[\textsc{model-selection scan}]~\textbf{Five open-weight VLMs, one protocol}: benign refusal (\%) on the borderline rung, text $\to$ blank canvas, no defense, no moderation layer, $100$ prompts, paired per prompt id, all cells in one job. ``disc.'' is discordant pairs (gained/lost); $p$ is exact McNemar. \texttt{qwen3-vl-8b} carries a hosted-scale effect on an open checkpoint, which is what bounds this paper's scope claim (\S\ref{sec:res-threshold}). Four of the five share one coarse alignment-tier label and span $+32$ to $+0$ points, so that label does not predict the effect.}
\label{tab:tierscan}
\end{table}

\section{Image properties, and instance replication}
\label{app:imgprops}

The main paper's property claim is carried by direct image-versus-image contrasts
(Table~\ref{tab:imgvsimg}), in which an image is attached in \emph{both} arms so that
only the named property moves. This section reports the wider sweep those contrasts
were drawn from, and the two axes for which we replicated a property level across
independently constructed files. The sweep is family \textbf{F2} of
\S\ref{app:multiplicity} and is controlled for false discovery rate, because the
question it asks is which arms move at all rather than whether any single one does;
the size replication is family \textbf{F6}.

Two readings depend on this sweep and are stated here rather than in the main paper.
\texttt{gpt-4o-mini} is property-invariant across every axis we varied, which is what
makes its null in Table~\ref{tab:imgvsimg} a measured null rather than an untested
one. And \texttt{gemini-2.5-flash} is null on every arm here, including the line
drawing at $+9$ points ($p=0.064$) and the caption at $+8$ points ($p=0.077$): the reason
the main paper withdraws the earlier universal claim and reports that model as
near-null in both collection windows.

A single row cannot separate a property from the particular file carrying it, since
each arm is one rendered file held byte-identical across its $100$ prompts. Two axes
are therefore replicated across independent instances: content, over three
independently constructed files per level at matched dimensions on three hosted
models (Table~\ref{tab:contentinstance}), and size, over three distinct fills at each
of four sizes on the open checkpoint (Table~\ref{tab:instance}). Colour is not
replicated and cannot be, because independent renders of a flat fill are
byte-identical.

\begin{table}[t]
\centering
\scriptsize
\setlength{\tabcolsep}{3pt}
\begin{tabular}{lcccc}
\toprule
 & \multicolumn{4}{c}{$\Delta$ benign refusal vs.\ that model's text arm (points)} \\
\cmidrule(lr){2-5}
Attached image & 4o-mini & fl.-lite & claude & flash \\
\midrule
\emph{text baseline} (\%) & \emph{13} & \emph{11} & \emph{11} & \emph{18} \\
\midrule
blank $512^2$ white       & \textbf{+31} & \textbf{+23} & \textbf{+38} & $-5$ \\
blank $256^2$ white       & \textbf{+32} & \textbf{+32} & --- & $0$ \\
blank $1536^2$ white      & \textbf{+33} & \textbf{+24} & --- & $-3$ \\
blank $1024\times141$ white & \textbf{+29} & \textbf{+27} & \textbf{+32} & $-6$ \\
blank $141\times1024$ white & \textbf{+30} & \textbf{+28} & --- & $-2$ \\
blank $512^2$ black       & \textbf{+27} & \textbf{+45} & --- & $+1$ \\
blank $512^2$ mid-grey    & \textbf{+33} & \textbf{+36} & --- & $0$ \\
line drawing              & \textbf{+31} & \textbf{+53} & --- & $+9$ \\
caption image \textsc{(ocr)} & \textbf{+30} & \textbf{+57} & \textbf{+60} & $+8$ \\
caption image, JPEG q40   & \textbf{+25} & \textbf{+56} & --- & $+1$ \\
\midrule
\emph{spread across arms} & \emph{8} & \emph{34} & \emph{28} & --- \\
\bottomrule
\end{tabular}
\caption{[\textsc{primary test}]~\textbf{Varying the image's properties one at a time.} Benign refusal shift vs.\ each model's own text arm, rung~2, no defense, $100$ prompts, paired per prompt id, all arms in one job. \textbf{Bold}: $p\leq1.5\times10^{-6}$, exact McNemar; every unbolded contrast is n.s. \texttt{gpt-4o-mini} is property-invariant ($8$ points across a sixfold resolution change, a transposed aspect ratio, three colours, two encodings and three content classes); \texttt{gemini-2.5-flash-lite} and \texttt{claude-sonnet-4-6} show a large content premium at \emph{matched} size ($+57$ vs $+27$, $+60$ vs $+32$ at $1024\times141$). \texttt{claude} was run on four load-bearing arms only, at $\sim$$20\times$ the per-cell cost of the others. All ten \texttt{gemini-2.5-flash} arms are null (\S\ref{sec:res-threshold}). \emph{Instances per property}: each row is \textbf{one} independently rendered file, held byte-identical across the $100$ prompts of its arm, so a single row cannot separate a property from the particular file carrying it. Two axes are replicated across independent instances and carry the property claims: size, over three distinct fills at each of four sizes (twelve arms, Table~\ref{tab:instance}), and content, over three independently constructed files per level at matched dimensions on three hosted models (Table~\ref{tab:contentinstance}). Colour on the hosted models is not replicated and cannot be: independent renders of a flat fill are byte-identical. Direct image-versus-image contrasts for this table are in Table~\ref{tab:imgvsimg}.}
\label{tab:imgprops}
\end{table}

\begin{table}[t]
\centering
\scriptsize
\setlength{\tabcolsep}{3.5pt}
\begin{tabular}{llcccc}
\toprule
model & level & blank & three instances & mean $\Delta$ & spread \\
\midrule
claude   & drawing $512^2$   & $47$ & $71$ / $67$ / $74$ & $\mathbf{+23.7}$ & $7.0$ \\
claude   & caption $1024{\times}141$ & $50$ & $76$ / $73$ / $70$ & $\mathbf{+23.0}$ & $6.0$ \\
fl.-lite & drawing $512^2$   & $34$ & $59$ / $63$ / $54$ & $\mathbf{+24.7}$ & $9.2$ \\
fl.-lite & caption $1024{\times}141$ & $38$ & $62$ / $71$ / $57$ & $\mathbf{+24.7}$ & $14.3$ \\
4o-mini  & drawing $512^2$   & $46$ & $37$ / $38$ / $39$ & $-8.0^{*}$ & $2.0$ \\
4o-mini  & caption $1024{\times}141$ & $43$ & $43$ / $43$ / $46$ & $+1.0$ & $3.0$ \\
\bottomrule
\end{tabular}
\caption{[\textsc{primary test}]~\textbf{Content levels replicated over independent instances} (benign borderline rung, no defense, $100$ prompts per arm, all thirty-three cells in one job, paired per prompt id against the size-matched blank re-run in that same job). Each \emph{level} is represented by three independently constructed files at identical dimensions (three clip-art line drawings, three caption sentences) so \emph{spread} is the between-instance range in refusal rate and is the quantity a single-file row cannot provide. \textbf{Bold}: all three instance contrasts significant at $p<0.001$, exact McNemar. $^{*}$: two of three at $p<0.05$ (the third $p=0.065$). The last row is n.s.\ throughout and is a null, not an unstable effect. Read the two right-hand columns together: wherever an effect exists, the spread across instances is a fraction of it, so the level and not the artifact is carrying it. The negative sign on \texttt{gpt-4o-mini} is against the \emph{blank}, not against text (its text arm refuses $13\%$; every image arm $37$--$46\%$). Colour is deliberately absent: independent renders of a flat fill are byte-identical, so colour admits no instance replication and is re-measured for run-to-run stability instead (discussed in text). Four cells lost $2$--$3$ rows to transient provider unavailability and are computed over $97$--$98$ pairs; the re-run threshold was fixed at $5\%$ before any result was read.}
\label{tab:contentinstance}
\end{table}

\begin{table}[t]
\centering
\small
\begin{tabular}{lcccc}
\toprule
& \multicolumn{4}{c}{$\Delta$ benign refusal (points) vs.\ text, by canvas size} \\
\cmidrule(lr){2-5}
fill & $256^2$ & $512^2$ & $1024^2$ & $1536^2$ \\
\midrule
white & +23 & +26 & +31 & \textbf{+36} \\
grey  & +20 & +33 & +36 & \textbf{+39} \\
black & +23 & +26 & +33 & \textbf{+35} \\
\bottomrule
\end{tabular}
\caption{[\textsc{replication}]~\textbf{Image-instance replication} on \texttt{qwen3-vl-8b}: four canvas sizes crossed with three fills, twelve independently rendered arms plus a shared text baseline ($51\%$), $100$ prompts each, one job, no defense, paired per prompt id. Every cell is significant at $p<0.001$ (exact McNemar against the text arm). The size trend is \emph{monotone in all three fills}, so the axis is a property rather than an artifact of one rendered file; the fill spread at fixed size ($3$--$7$ points) stays well inside the size span ($12$--$19$ points), replicating colour-inertness at four sizes. Each arm is byte-identical within itself and all twelve differ from one another, verified by hash (\S\ref{sec:res-threshold}).}
\label{tab:instance}
\end{table}

\section{Within-family replication}
\label{app:generational}

Three consecutive checkpoints of one family, with family, size class, serving stack, judge and the rendered inputs themselves all held fixed so that only the checkpoint moves. This is the tightest control under which the main paper's decoupling claim can be stated: the benign cost runs $+8$, $+1$ and $+28$ points while the harmful effect stays inside $\pm10$ points at every rung. We describe the benign pattern as a step between adjacent checkpoints of \emph{this} family rather than as the date the behaviour entered the world, and we do not attribute it to post-training: a release changes alignment, vision encoder and pretraining corpus together.

\begin{table}[t]
\centering
\scriptsize
\setlength{\tabcolsep}{3pt}
\begin{tabular}{lccccccc}
\toprule
 & \multicolumn{4}{c}{benign refusal (\%)} & \multicolumn{3}{c}{harmful ASR (\%)} \\
\cmidrule(lr){2-5}\cmidrule(lr){6-8}
Generation & text & blank & $\Delta$ & $p$ & text & blank & $\Delta$ \\
\midrule
qwen2-vl-7b \emph{(oldest)}   & 72$^{\dagger}$ & 80 & +8 & 0.057 & 2 & 2 & \phantom{$-$}0 \\
qwen2.5-vl-7b                 & 37 & 38 & +1 & 1.000 & 4 & 7 & +3 \\
\textbf{qwen3-vl-8b} \emph{(newest)} & 54 & \textbf{82} & \textbf{+28} & $\mathbf{<10^{-4}}$ & 2 & 1 & $-$1 \\
\bottomrule
\end{tabular}
\caption{[\textsc{exploratory follow-up}]~\textbf{One family, three generations, both sides}: text $\to$ blank canvas, no defense, same canonical text, paired per prompt id, $n=100$ per arm. Family, size class, serving stack, judge (\texttt{gpt-5-mini}) and the rendered inputs themselves are held fixed; only the checkpoint moves. Benign: the effect is a \emph{step} at the newest rung, not a gradient. We describe this as a step between adjacent checkpoints of \emph{this} family, not as the date the behaviour entered the world: we observe three shipped checkpoints of one family, which cannot say when or whether it appeared elsewhere. A release changes alignment, vision encoder and pretraining corpus together, so this does \emph{not} isolate post-training. Harmful: every rung is a bounded null, with $95\%$ intervals $[-4.5,+4.5]$, $[-2.7,+9.3]$ and $[-6.1,+3.7]$ points, all equivalent to zero within a $\pm10$-point margin and the oldest within $\pm5$ points. The two sides do not move together at any rung. $^{\dagger}$\,\texttt{qwen2-vl} refuses $72\%$ with no image at all, so its $+8$ points is measured against a partly saturated baseline and is not magnitude-comparable to the newest rung (\S\ref{sec:res-threshold}).}
\label{tab:generational}
\end{table}

\section{Break-even sensitivity}
\label{app:breakeven}

Arithmetic on the measured rates only; no new experiment. The main paper reports the benign cost and the prevented harm side by side and never as a ratio, because a ratio is dominated by a denominator that is vanishing. For a reader who nonetheless wants the exchange rate written down, this is how many times worse one harmful completion must be than one false refusal for the canvas to pay for itself. The model-dependent term spans a factor of $2.4$ across three models while the prevalence term spans a factor of $100$ over the range shown, so the trade is governed mainly by a quantity we do not measure. Severity, partial compliance, and the availability of other routes to the same answer are not modelled and would each shift the threshold further.

\begin{table}[t]
\centering
\scriptsize
\setlength{\tabcolsep}{4pt}
\begin{tabular}{lcccc}
\toprule
 & & \multicolumn{3}{c}{break-even $k^{\star}$ at prevalence $\pi$} \\
\cmidrule(lr){3-5}
Model & $\Delta_{\text{ben}}/\Delta_{\text{harm}}$ & $\pi{=}10\%$ & $\pi{=}1\%$ & $\pi{=}0.1\%$ \\
\midrule
claude-sonnet-4-6     & $2.8$ & $26$ & $280$ & $2{,}800$ \\
gpt-4o-mini           & $6.8$ & $61$ & $670$ & $6{,}800$ \\
gemini-2.5-flash-lite & $3.8$ & $34$ & $380$ & $3{,}800$ \\
\bottomrule
\end{tabular}
\caption{[\textsc{sensitivity analysis, no new data}]~\textbf{How many times worse must one harmful completion be than one false refusal for the canvas to pay for itself.} Arithmetic on the measured borderline-rung rates only ($\Delta_{\text{benign}}=+51,+34,+23$ points against $\Delta_{\text{harmful}}=18,5,6$ points of prevented harm); no new experiment. $k^{\star}=\frac{1-\pi}{\pi}\cdot\frac{\Delta_{\text{benign}}}{\Delta_{\text{harmful}}}$, rounded to two significant figures. The model-dependent term spans a factor of $2.4$ across these three models, while the prevalence term spans a factor of $100$ over the range shown, so the trade is governed mainly by a quantity we do not measure, which is precisely why we report the two sides separately and never as a ratio. Severity, partial compliance, and the availability of other routes to the same answer are not modelled and would each shift $k^{\star}$ further.}
\label{tab:breakeven}
\end{table}

\section{Pairing protocol, and what each arm varies}
\label{app:elab4}

This section gives the dispatch protocol used for the paired contrasts, and states
which quantity each family of arms varies.

\subsection*{Pairing protocol.}
\label{app:fparagraphpairin}

Both arms of a pair are dispatched concurrently under one shared limit; across
the $127$ paired arms in our records every pair starts within the same second and the
windows overlap by a median of $77\%$. Because concurrent dispatch bounds an ordering
confound rather than excluding it, we also re-collected the central contrast under
\emph{randomised paired interleaving}, where the two conditions are never separately
dispatched. The two designs agree on all three targets tested; the paired comparison, its shuffle seeds and its two caveats are in \S\ref{app:interleave}.

\subsection*{What is varied.}
\label{app:fparagraphwhatis}

\textbf{Image properties}: direct contrasts between
two attached-image arms differing in one property (colour, rendered content, or pixel
count) over one canonical text. \textbf{Serving route}: one open checkpoint reached
through a managed host and through our own vLLM, holding weights and post-training fixed
by construction. \textbf{Mention}: a $2\times2$ crossing attachment with a
system-message mention of an image, worded so the identical string is true whether or not
an image is attached.

\section{The request-independent images}
\label{app:images}

\section{The result in full, and the controls behind it}
\label{app:elab3}

\subsection*{The conclusion, in full.}
\label{app:fsectionconclusi}

The effect this paper measures reaches the model through the image channel. On one checkpoint a second
channel carries the same shape, since asserting an attachment that does not exist moves
refusal without the word ``image'' contributing any of it. The behaviour needs no serving stack and is not a property of vision--language models as
such; it belongs to particular aligned checkpoints, hosted or open, with the caveat that
on the hosted models we show only that a stack is not \emph{necessary}. Three
open-weight models give nothing and a fourth we serve ourselves gives a hosted-scale
$+32$ points, and no label we tried predicts which. One caveat belongs here rather than
only in the Supplementary Document: \texttt{qwen3-vl-8b} was \emph{selected} from a
five-model scan for its large effect, so the generational step, the token-budget control
and the factorial are follow-ups on a selected checkpoint. Its presence effect is
independently re-collected three times, so the effect is not a selection artifact, but
the finer structure read off it is single-family evidence. What makes this an alignment problem rather than a usability one is the price and who
pays it. The cost lands on sensitivity-adjacent benign traffic, meaning benign questions
about privacy, self-harm, violence and illegal activity, and it is decoupled from what it
buys: on the borderline rung the canvas costs $+51$, $+34$ and $+23$ points of benign
refusal against $18$, $5$ and $6$ points of prevented harm, and on a checkpoint already
yielding a harmful completion on only $2\%$ of plain harmful requests it still costs $29$
points with nothing computable to set against it. As models get better at refusing
harmful text the denominator shrinks and the cue increasingly charges for nothing. Whether the trade is \emph{net} negative in a given deployment we do not claim, since it
turns on a prevalence and a relative severity we never measure
(\S\ref{app:breakeven}). Its sign is not even fixed: on one open model the same canvas
raises attack success $33$ points. Image presence is not a conservative default that a
deployer chose and priced. It is an uncontrolled variable, one an open checkpoint carries
with no serving layer anywhere in the path, and the first step toward controlling it is
measuring what it costs.

\subsection*{Same weights, two serving stacks.}
\label{app:fparagraphsamewe}

Because a non-significant test on so few discordant pairs is
consistent with almost anything, we report the route difference as an interval rather
than a failure to reject: $[-2.5,+5.1]$, $[-2.2,+4.5]$, $[-2.2,+4.5]$ and $[-8.0,+1.0]$
points, so every arm is equivalent within a $\pm10$-point margin and both blank arms
within $\pm5$ (Table~\ref{tab:sameweights}). Two limits keep this from settling the larger question. \texttt{gemma-3-12b-it} is a
small-effect model, $+7$ points where the hosted models give $+23$ to $+54$, so route
equivalence is established where there is not much to differ about, and one checkpoint is
not a survey. A larger-effect version of the control is available in principle and
\S\ref{app:routecontrol} says why we do not report it: on the checkpoint we would want,
magnitude tracks the visual-token count, which the serving stack's preprocessing decides
and a managed endpoint does not expose, so a naive route comparison would vary the input
and the route together. What the arm does rule out is the narrow deflationary reading,
that the shift measured on hosted APIs is manufactured by serving infrastructure rather
than by the model. It does not establish the converse: the hosted shifts are measured
through stacks we cannot open, and how much those contribute stays unresolved.

\subsection*{The cost is not spread evenly, which is what makes it an alignment cost.}
\label{app:fparagraphthecos}

They are moving a decision boundary, and the traffic sitting against that boundary is benign questions about privacy, self-harm, violence and illegal activity, asked by people entitled to answers. Re-measured on a category-stratified sample across all ten OR-Bench categories, all thirty per-category contrasts move in the same direction, $26$ survive a Benjamini--Hochberg correction, and the cost is \emph{larger} than the first slice showed.

\subsection*{The benefit side, on a matched population.}
\label{app:fparagraphtheben}

ASR falls: \texttt{claude-sonnet-4-6} $18\!\to\!0$ ($p=7.6\times10^{-6}$), \texttt{gemini-2.5-flash-lite} $11\!\to\!5$ ($p=0.031$), \texttt{gpt-4o-mini} $6\!\to\!1$ (n.s.), \texttt{gemini-2.5-flash} $14\!\to\!13$ (n.s.). The shift is therefore real in both directions, and the paper's claim is not that image presence fails to help.

\subsection*{Relation to safety-perception distortion.}
\label{app:fparagraphrelati}

The substantive gap is on the benign side: their table reports the effect of \emph{their proposed correction} on false alarms rather than the effect of image presence itself, so the cost this paper prices is not quantified there. \S\ref{app:novelty} sets the two out item by item.

\section{What moves the threshold, and what it costs}
\label{app:elab2}

\subsection*{The response varies with properties that carry no risk information.}
\label{app:eparagraphtheres}

On \texttt{gemini-2.5-flash-lite} a black canvas costs $+22$ points more than a white one of identical size ($22$ prompts flip against $0$, $p=4.8\times10^{-7}$), and a rendered caption $+30$ points more than a size-matched blank; \texttt{claude-sonnet-4-6} shows the same content premium ($+28$ points, $p=2.5\times10^{-7}$). On the open checkpoint the price climbs with canvas size: $256^2$ against $1536^2$ is $+16$ points, $18$ flips against $2$ ($p=0.0004$). And on \texttt{gpt-4o-mini} the identical pair of arms differs by $+1$ points ($p=1.00$), that model is property-invariant, holding inside an $8$-point band across a sixfold resolution change, a transposed aspect ratio, three colours, two encodings and three content classes (\S\ref{app:imgprops}). So attachment is what triggers the shift, and the image decides how much it costs, by an amount that is itself a property of the model. \textbf{What carries that magnitude is outside this paper's claim.} Attaching an image is not a single switch: it changes how the request is serialised, introduces modality tokens, routes the input through a vision encoder, and applies the provider's own preprocessing. A black-box experiment moves all four together, so what we measure is the effect of that whole interface, and we call it the \textbf{attachment-interface effect} rather than implying a lone semantic bit. Nothing in the argument requires the four to be separated: a deployer changes them together too, and the claim here is about the response, not its mechanism. We take this as strengthening rather than weakening the claim. A safety-relevant cue that fires on an empty canvas is already hard to defend; a cue whose magnitude then swings on choices as incidental as canvas colour and pixel count, differently on each model, is not a conservative default anyone priced: it is the uncontrolled variable of our title.

\subsection*{What the instruction cue is made of.}
\label{app:eparagraphwhatth}

Adding the attachment clause, any \emph{file} attached is a fixed placeholder and should be disregarded, costs a further $+16$ points ($p=0.0015$). Replacing \emph{file} with \emph{image}, the single word the deflationary reading turns on, is worth $\mathbf{-1}$\textbf{ points} ($12$ gained, $13$ lost, $p=1.0$; at $25$ discordant pairs this excludes an image-word effect above $\sim\!11$ points). The flash effect is therefore real, not visual, and not lexically about images: what moves that model is being told an attachment exists and should be ignored, while nothing is attached. That is the same shape as the presence effect measured everywhere else in this paper (a property of the request's form, carrying no per-prompt information about what is asked, moving the refusal threshold) reaching the model through the text channel instead of the image one. It also disciplines how the instruction arm may be read. On the three models above it is a mitigation that mostly fails; on \texttt{gemini-2.5-flash} it is a second manipulation, and the $+33$-point cell must not be reported as an image effect. We give the instruction results per model for that reason and never pool them.

\subsection*{It is attachment, not the mention of an image.}
\label{app:eparagraphitisat}

With the mention text held byte-identical across both arms, attachment still moves refusal $\mathbf{+20}$\textbf{ points}: $20$ prompts gained a refusal and \textbf{not one} moved the other way ($p=1.9\times10^{-6}$). Mention alone gives $+8$ points and does not reach significance. Attachment is the operative variable on this checkpoint. Two limits, stated so the table is not read for more than it shows. The mention cell is \emph{underpowered rather than null}: at the $16$ discordant pairs observed, the exact test could only have resolved a shift of about $10$ points, and we measured $8$. We can exclude a mention effect the size attachment produces, not a small one. And the apparent $-10$ points interaction is the difference between two individually non-significant contrasts, so we do not report an interaction. The \texttt{gemini-2.5-flash} configuration that motivated the test is hosted and remains open.

\subsection*{The cost is spread across topics, and the first measurement understated it.}
\label{app:eparagraphthecos}

The effect is broad. \textbf{All thirty per-category
contrasts move in the same direction}, and $26$ survive a Benjamini--Hochberg
correction across the family. It is also \emph{larger} than the unrepresentative
slice reported: $+52.3$, $+36.3$ and $+34.7$ points against the $+51$, $+34$ and $+23$ points
of Table~\ref{tab:ladder}, so the original slice was conservative and correcting it
raises the measured cost rather than lowering it. Which topic pays most is strongly
model-specific and shares no ordering across models: \texttt{claude-sonnet-4-6} is
worst on \emph{deception} ($6.7\%\!\to\!86.7\%$), \texttt{gpt-4o-mini} on
\emph{harmful} ($+53$ points) and \texttt{gemini-2.5-flash-lite} on \emph{sexual}
($+57$ points). This is the same checkpoint-specificity the property sweep and the
sign reversal show on their own axes.

\subsection*{It is presence, not that particular image.}
\label{app:eparagraphitispr}

Both request-independent image classes, the blank canvas and the line drawing, inflate refusal on \texttt{claude-sonnet-4-6} ($+54$\,/\,$+48$ points), \texttt{gpt-4o-mini} ($+30$\,/\,$+31$ points) and \texttt{gemini-2.5-flash-lite} ($+23$\,/\,$+52$ points), every contrast at $p\leq1.5\times10^{-6}$. The two classes share no content and differ in size, aspect ratio and whether anything is drawn at all; the operative factor is presence. They also share one property that matters for the obvious alternative explanation: \emph{neither contains any readable text}, so no OCR path exists in either arm. The one request-independent image we test that \emph{is} OCR-able, a caption rendered on white, inflates benign refusal too, so the effect appears on both sides of that line.

\subsection*{What it costs, what it buys, and why those are not the same question.}
\label{app:eparagraphwhatit}

The cost does not depend on there being anything to buy. \texttt{qwen3-vl-8b} yields a harmful completion on only $2\%$ of plain harmful requests, below every hosted model we measure, so the canvas has almost nothing left to prevent there; the benign side still moves $+29$ points. Across our models the harmful-side denominator collapses as alignment improves while the benign cost does not track it down. A cue weighed against $18\%$ residual attack success is a different proposition from the same cue weighed against $2\%$, and the direction of travel is toward the second. That is also why a single ratio would mislead: it is dominated by a denominator that is vanishing. Nor is the sign fixed. On \texttt{pixtral-12b} the identical blank canvas raises attack success $48\!\to\!81\%$ ($p=1.0\times10^{-7}$), stripping the model's safety hedging rather than changing what it knows.

\subsection*{The cost side scales with topical sensitivity.}
\label{app:eparagraphthecos-b}

This is a \emph{threshold shift}, not blanket caution: the cost falls not on benign traffic in general but specifically on sensitivity-adjacent benign traffic. The ten OR-Bench categories span deception, harassment, harmful advice, hate, illegal activity, privacy, self-harm, sexual content, unethical conduct and violence. The neutral rung doubles as a validity gate, and it passes: no-image baselines sit at $0$--$2\%$.

\subsection*{The open checkpoint's cost is measured three times.}
\label{app:eparagraphtheope}

Spot-checking the raw rows before believing it: zero empty responses and zero judge errors on every arm, and the flips are unambiguous. With the same prompt and the same blank white canvas attached, the text arm engages (``It's important to clarify that while some techniques\ldots'') where the canvas arm opens ``I can't help with that.''

\subsection*{On the open checkpoint the benign cost is paid against a floored harmful side.}
\label{app:eparagraphontheo}

\textbf{The benign side still moves $29$ points.}

We are careful about what that does and does not show. The harmful contrast is a $1$-versus-$2$ prompt difference: an \emph{underpowered null}, not a demonstrated zero. A null measured at a floor cannot tell us what the cue would do where there is in fact something to prevent, so rather than argue from it we removed the floor and measured.

\subsection*{The instruction does not neutralise it.}
\label{app:eparagraphtheins}

It does not work. The shift falls from $+54$ to $+42$ points on \texttt{claude-sonnet-4-6} ($p=0.0018$), $+30$ to $+23$ points on \texttt{gpt-4o-mini} ($p=0.039$) and $+23$ to $+17$ points on \texttt{gemini-2.5-flash-lite} (n.s.): \textbf{roughly three-quarters of the effect survives an explicit instruction to ignore the image}. \texttt{gemini-2.5-flash} inverts it. The blank canvas alone does nothing to that model ($-1$ points, n.s.), but the same canvas \emph{plus} an instruction to disregard it drives benign refusal from $15\%$ to $47\%$ ($+32$ points, $p=1\times10^{-8}$). On that model the mitigation is the manipulation.

\subsection*{Taken together.}
\label{app:eparagraphtakent}

The same-weights arm constrains where that variability can live: the hosting layer is inert on the one checkpoint we can test both ways, and among the open checkpoints we serve ourselves, where the serving arrangement is fixed by construction, the behaviour still ranges from $+32$ points to a sign reversal, so the checkpoint alone is enough to produce it. What the hosted stacks add to their own larger shifts is not identified by this design. A feature whose \emph{sign} depends on which aligned model is behind the endpoint is not a conservative default that costs some utility; it is an uncontrolled variable that happens to point the safe way on the models most people evaluate.

\section{Why this cost goes unmeasured}
\label{app:elab5}

Isolating the effect requires an image that cannot carry content, which is the control this paper is built on. We attach a \emph{blank canvas} to a benign request, an image that cannot be OCR'd, cannot relate to the request, and is byte-identical across every prompt in its condition, so it carries literally zero per-prompt information. Refusal rises by tens of points. Nothing in this paper places a defense in the loop, so the effect is a property of the models themselves rather than of anything wrapped around them.

We show that aligned VLMs condition their refusal threshold on attachment anyway:
substantially, reproducibly, and on models a deployer can download and inspect.

\section{Mechanism, setup, and variation across checkpoints}
\label{app:elab6}

\subsection*{What we can and cannot say about the mechanism.}
\label{app:kparagraphwhatw}

The black-box limit on identifying a mechanism is real, but it is not the whole picture: on the one checkpoint we serve ourselves, weight access lets us hold three of the four things attachment changes fixed and vary the fourth, which excludes provider preprocessing, total input length, serialisation and encoder invocation as explanations of a magnitude that \emph{varies} across those arms. Which property of the visual input then carries it is a question we pose and deliberately do not answer here. Three further observations constrain the rest. It is not image \emph{content}: two images sharing nothing reproduce it, and one is a uniform blank canvas that cannot be OCR'd or related to any request. It is not exclusively perceptual either: on \texttt{gemini-2.5-flash} a system message asserting an attached image raises benign refusal $+32$ points while the image alone does nothing. The placebo ladder locates that effect precisely (Table~\ref{tab:placebo}): it is not the modality word ($-1$ points, n.s.) and it does not need an image to be present, so what that checkpoint keys on is the assertion that an attachment exists. Two channels, one shape: in both cases a property of the request's form, carrying no \emph{per-prompt} information about what the request asks, moves the refusal threshold. Which channel a given checkpoint responds to is not something we can predict. The picture is awkward: content is frequently ignored, yet presence moves the threshold by tens of points. The third observation constrains where the variability can live. A convenience sample of three open-weight models would license reading the cost as a property of a moderated serving stack; the serving-route control (Table~\ref{tab:sameweights}) and the open, self-served \texttt{qwen3-vl-8b} at $+32$ points (Table~\ref{tab:tierscan}) both refuse that reading, and on \texttt{pixtral-12b} the same manipulation changes sign outright. What survives is that image presence perturbs the safety threshold, and that among the seven checkpoints we serve ourselves, under one serving arrangement throughout, which way it moves is set by the checkpoint. Whether the same holds behind the hosted stacks we cannot open is not something this design can say, and which component of the checkpoint carries it (post-training, vision encoder, pretraining corpus) a black-box comparison of shipped models cannot say either.

\subsection*{Experimental setup, in full.}
\label{app:ksectionexperim}

Exact identifiers, decoding settings, serving versions and verbatim system messages are in \S\ref{app:repro}. \textbf{Harmful prompt set:} OR-Bench's harmful split: the constructed counterpart of rung~2, same pipeline and topic distribution, which is what makes the two sides of the price comparable. Each cell uses $100$ prompts. \textbf{Benign refusal rate}: the fraction of $100$ benign prompts whose response is judged a refusal by the JailbreakBench refusal classifier; \textbf{lower is more useful}. Both judges run on \texttt{gpt-5-mini} throughout, and the two rubrics are never substituted for one another: scoring a harmful arm with the refusal rubric would silently measure a different quantity. Empty responses are auto-classified as refusals, which handles upstream API filtering correctly. The collection window, configuration, judge and protocol for \emph{every} result in this paper are in Table~\ref{tab:provenance}.

\subsection*{Where this sits among multimodal safety findings.}
\label{app:kparagraphwhere}

Every one of these varies either what the request asks or what the image contains. None
holds the request fixed and varies only \emph{whether} an image is attached, which is the
axis this paper isolates and prices on both sides.

\subsection*{The behaviour arrives as a step, not a gradient.}
\label{app:kparagraphthebe}

The ordering does not climb with post-training recency; it jumps at the newest rung. What that supports is a discontinuity between two consecutive checkpoints of this family rather than a property accumulating with scale: not a dose of alignment. We are describing where the step falls in one family's release order, not claiming the behaviour originated there; other families may have acquired it at other points, or never. It does not identify \emph{which} change: consecutive releases in this family differ in vision encoder, tokenisation and pretraining corpus as well as in alignment, and three shipped checkpoints cannot separate them. What we claim is narrower and still load-bearing: the behaviour is not a smooth function of generation, and it is carried by something a downloader receives with the weights. One caveat belongs with the table rather than beneath it: \texttt{qwen2-vl-7b} refuses $72\%$ of these benign prompts with no image attached at all, leaving only $28$ points of headroom, so its $+8$ points is measured against a partly saturated baseline and is not magnitude-comparable to \texttt{qwen3-vl}'s $+28$ from a $54\%$ base.

\subsection*{The sign is not fixed either.}
\label{app:kparagraphthesi}

Under our manipulation attack success on \texttt{llava-1.5-7b} moves $37\!\to\!76\%$
($+39$ points, $p=3.6\times10^{-12}$), and the paired structure is as clean as any
contrast here: of $39$ discordant prompts, $39$ gained a harmful completion under the
canvas and none lost one. That checkpoint inverts on the benign axis too, refusal falling
$10\!\to\!2\%$, though at $1/9$ discordant pairs that movement does not survive the
family-wise correction of \S\ref{app:multiplicity} and we rest nothing on it. With two of
seven open checkpoints inverting, and both weakly aligned, we establish that the direction
varies, not what fixes it. Inspecting responses shows the shift is not a judge artifact:
on a phishing request the text arm opens ``this is illegal and unethical\ldots severe
legal consequences'' before answering while the blank-canvas arm opens ``Here is a
detailed guide:''. The canvas strips the safety framing rather than changing what the
model knows.

\section{Ethics statement, in full}
\label{app:ethics}

All experiments use standardized public prompt benchmarks (HarmBench, JailbreakBench, OR-Bench) released for safety evaluation. \textbf{We introduce no attack.} The only manipulation studied here is attaching a request-independent image (a blank canvas, a line drawing, or a rendered caption) to an otherwise unchanged request; we release no adversarial assets, and nothing in this paper increases what an attacker can do that attaching an arbitrary image did not already permit. The finding does describe a cue that an attacker aware of it could exploit, and we judge disclosure to be net-positive: the cue is trivially discoverable by anyone who attaches an image, the affected checkpoints are public, and the cost we document falls on benign users (people asking benign questions about privacy, self-harm, violence and illegal activity) who cannot discover or avoid it without being told it exists. Appendix examples deliberately use prompts whose harm is informational rather than physically actionable.

\section{Elaboration of results and discussion}
\label{app:elab}

Each subsection continues a paragraph of the main paper, which states the claim and its numbers.

\subsection*{The decoupling holds inside one family; the \emph{trend} cannot be tested there.}
\label{app:paragraphthedeco}

Within the one family we can step through, the two sides come apart. At the newest rung the canvas costs $+28$ points of benign refusal
($p=2.5\times10^{-7}$) bought against a harmful effect confined to
$[-6.1,+3.7]$ points, with byte-identical rendered inputs on both sides. That is the
controlled form of the claim, and it is the one we rest on. The same table refuses to support a stronger reading, and we flag it rather than let
the reader supply it. This family cannot test whether the harmful side \emph{collapses}
with recency, because it never had the headroom: plain-text attack success is $2$, $4$
and $2\%$, already at the floor at the \emph{oldest} rung. A flat harmful line across a
ladder that starts floored is uninformative about a downward trend rather than evidence
against one. It does establish a smaller thing worth stating: \texttt{qwen3-vl}'s
$2\%$ is not a recency effect within its own family, since its two-generation-older
sibling sits at $2\%$ as well. The collapse therefore remains an across-model
observation, carried by the hosted models that genuinely span $18$ to $6\%$, and we do
not claim the denominator shrinks with recency \emph{within} a family. The claim is the asymmetry visible across the whole set: the harmful side collapses across the models we measure while the benign cost does not track it down. A cue priced against $18\%$ residual attack success is a different proposition from the same cue priced against $2\%$, and the direction of travel is toward the second. This is why we report the decoupling as the finding, and report the two sides directly rather than reducing them to a ratio that would be undefined here.

\subsection*{Given headroom, the cue does buy something, but not in a fixed direction.}
\label{app:paragraphgivenhe}

A cue that suppresses a plainly-worded harmful request, suppresses a set-theoretic one, and does neither to a code-formatted one is responding to the surface form of the input rather than to what the input asks for: the same failure of control the benign side shows, now visible on the harmful side. On the plain arm the effect is entirely one-sided ($10$ prompts lost, $0$ gained), and reading those ten shows a mechanism narrower than a threshold shift: they are responses that refuse in their first sentence and then supply the payload regardless, which the canvas converts into refusals that hold. One scope limit belongs here rather than in a footnote. This experiment necessarily leaves the OR-Bench harmful split, that split \emph{is} the floor, so it establishes how the cue behaves on harmful requests with headroom, and it is not a matched exchange against the benign rung of Table~\ref{tab:owladder}.

\subsection*{But it does not require a hosted serving stack either, and a sample of open-weight nulls is what makes that easy to miss.}
\label{app:paragraphbutitdo}

We set the two arms out in that order because the inference they invite, generalising from a convenience sample of open models to a claim about model classes, is one this literature makes routinely, and the scan is what forecloses it. The scan also refutes the obvious ordering hypothesis. Four of the five models carry the same coarse ``alignment tier'' label and span $+32$ to $+0$ points; the within-label spread dwarfs any between-label gap, so that label does not predict the effect and we report it as a null rather than dressing it as a dose--response. One boundary belongs with that null rather than beneath it: the scan spans the middle and weak tiers only. The single strongly-aligned open vision--language model available to us cannot be served at all (support for its architecture was withdrawn from the inference engine we use, and no version pairing of that engine with a compatible transformers release restores it) so the ordering is tested across two tier levels, not three. A strongly-aligned open checkpoint would be the sharpest test of the hypothesis and we could not run it.

\subsection*{Shortcut-driven safety judgements.}
\label{app:paragraphshortcu}

The shortcut \citet{hinojosa2026saves} identify is
about what the image \emph{depicts}: a symbol inside the image carries a danger
prior, and every image in that design is load-bearing by construction, since
safety cannot be determined from the instruction alone. Ours is about whether an
image \emph{exists}, with a canvas that depicts nothing and a request whose
harmfulness is fully determined by its text. Theirs is a content-level shortcut
and ours is a form-level one; a design in which the image decides the answer
cannot reach the limiting case we measure. That the two lines meet on
\texttt{qwen3-vl-8b} (the open checkpoint carrying our largest open-weight
effect) is corroboration from an independent manipulation rather than a
duplicate finding.

\subsection*{What the open checkpoint excludes, and what we leave open.}
\label{app:paragraphwhatthe}

Provider preprocessing is absent by construction, since we serve the weights with no vendor wrapper, and the same-weights route control of Table~\ref{tab:sameweights} bounds any inference-implementation contribution separately. Total input length is excluded by the padded text ladder, which buys $0$ points at the largest canvas's own token budget against that canvas's $+32$. Serialisation and encoder invocation are \emph{binary}: every canvas arm attaches exactly one image, so the request is serialised the same way and the vision encoder runs in all of them, and neither can therefore explain a magnitude that \emph{varies} across those arms. What remains is the visual input itself, and we stop there deliberately. Identifying which of its properties carries the magnitude is a different question from the one this paper asks, it requires an architecture-aware design that the hosted tier cannot support (a canvas can only buy visual tokens on an encoder whose output length grows with the image, and most deployed stacks pin every image to a fixed budget) and we report it separately rather than half-answering it here. What the arms above do establish is what the claim needs: the response is not invariant to variation carrying no information about the request.

\subsection*{What the two sides do and do not license.}
\label{app:paragraphwhatthe-b}

The first is a property of the traffic, and we do not measure it. Table~\ref{tab:breakeven} evaluates $k^{\star}$ over three prevalences, and the sensitivity is severe enough to be the point: at $1\%$ harmful traffic a harmful completion must be $280$ to $670$ times worse than a false refusal before the canvas breaks even, and at $0.1\%$ the requirement is thousands-fold. We draw no welfare conclusion from this. The table exists to make explicit that the sign of the trade is set by an assumption the paper does not measure, and that our two measured rates alone cannot settle it. Three further quantities we do not model would each move it: harm severity is not binary, refusals and completions are not all-or-nothing (partial compliance sits between them), and a refused user often has other routes to the same answer. These are calibration measurements of how a model's threshold responds to request form, not an estimate of a net safety cost.

\subsection*{Scope, and what we do not claim.}
\label{app:paragraphscopean}

We do not claim image presence never helps, or that a threshold shift is illegitimate: the harmful-side reduction is real on models with headroom, and a deployer who values a prevented completion above several benign refusals may rationally accept the trade. We claim the cost should be known before it is accepted, that it is paid against a denominator that shrinks as models improve, and that a safety-relevant behaviour whose magnitude swings with canvas colour and whose \emph{sign} swings with the checkpoint is not a default anyone designed. Finally, we do not identify \emph{which} property of the visual input carries the magnitude. We exclude several candidates (\S\ref{sec:res-threshold}) and leave the positive question open, because answering it requires an architecture-aware design the hosted tier cannot support.

\subsection*{The claim, and what would refute it.}
\label{app:paragraphtheclai}

A safety-relevant prior should be invariant to variables that carry no information about what is being asked. That invariance is directly testable, and it fails on four independent measurements. \emph{It is a function of variables carrying no risk information.} A black canvas costs $+22$ points more than a white one of identical size ($22$ prompts flip against $0$, $p=4.8\times10^{-7}$), and on an open checkpoint the price climbs with pixel count ($+16$ points from $256^2$ to $1536^2$, $18$ flips against $2$, $p=0.0004$). \emph{Its direction is not shared}: the identical canvas raises attack success on two open checkpoints while tightening refusal on four hosted ones. \emph{It does not update on evidence}: an instruction stating the image is a placeholder to be disregarded removes only $6$ to $12$ points of a $23$ to $54$-point shift. \emph{And it does not require an image at all}: on one model a sentence asserting an attachment exists moves refusal $+16$ points with nothing attached, the word ``image'' contributing none of it. A prior conditions on the informative variable; this conditions on canvas colour, pixel count, and the assertion. Note where the evidence falls, because it blocks the obvious dismissal: the nuisance-variable dependence is measured on the frontier hosted models and the sign inversion on open weak ones, so neither tier can be set aside as the other's failure. We therefore describe it as a \emph{shortcut}, and shortcuts fail two ways, \emph{steerable} by anyone who knows the key and \emph{mis-generalising} to everything the key does not track. This paper measures both.

\section{Tables supporting claims stated in the main text}
\label{app:tables}

\begin{table}[t]
\centering
\small
\begin{tabular}{lcc}
\toprule
added input & \multicolumn{2}{c}{benign refusal \% ($\Delta$ vs.\ text baseline)} \\
\cmidrule(lr){2-3}
tokens & as text filler & as blank canvas \\
\midrule
$\approx\!64$    & $47$ ($-7$, n.s.)  & $73$ (\textbf{$+19$}) \\
$\approx\!260$   & $50$ ($-4$, n.s.)  & $76$ (\textbf{$+22$}) \\
$\approx\!2310$  & $54$ ($\pm0$, $p=1.00$) & $86$ (\textbf{$+32$}) \\
\bottomrule
\end{tabular}
\caption{[\textsc{primary test, control arm}]~\textbf{Input length is not what the model charges for} (\texttt{qwen3-vl-8b}, benign borderline rung, no defense, $100$ prompts, all seven arms in one job, paired per prompt id, shared text baseline $54\%$). Each row holds the added input budget fixed and changes only whether it arrives as text or as a blank white canvas. Budgets are measured on the target's own accounting ($63$/$273$/$2310$ text against $66$/$258$/$2306$ visual), so the largest text arm carries $4$ more tokens than the image arm it is matched against. Text padding never exceeds the baseline at any budget, while the canvas raises refusal at every budget; at the matched extreme the two are $32$ points apart ($33$ flips to $1$, $p=4.1\times10^{-9}$). The canvas column re-measures three white arms from an earlier job in a fresh target and judge pass and agrees within $1$ points absolute.}
\label{tab:tokenbudget}
\end{table}

\begin{table}[t]
\centering
\scriptsize
\setlength{\tabcolsep}{3pt}
\begin{tabular}{lccrrr}
\toprule
paired contrast & refusal \% & $\Delta$ & gained & lost & $p$ \\
\midrule
mention, no image        & $51\!\to\!59$ & $+8$           & $12$ & $4$ & $0.077$ \\
attachment, no mention   & $51\!\to\!81$ & $\mathbf{+30}$ & $32$ & $2$ & $6.9{\times}10^{-8}$ \\
attachment, mention held & $59\!\to\!79$ & $\mathbf{+20}$ & $20$ & $0$ & $1.9{\times}10^{-6}$ \\
\bottomrule
\end{tabular}
\caption{[\textsc{primary test}]~\textbf{Attachment $\times$ mention} on \texttt{qwen3-vl-8b}, benign refusal (\%) on the borderline rung, no defense, $100$ prompts, paired per prompt id, all four cells in one job. The four factorial cells are refusal $51$ (no image, no mention), $59$ (no image, mention), $81$ (image, no mention) and $79$ (image, mention); every row below is one paired contrast between two of them and is self-contained, so its $\Delta$, counts and $p$ can be checked without reference to any other row. The mention is a system message worded so the identical string is true whether or not an image is attached, so the mention factor is byte-identical across attachment conditions. \emph{Gained}/\emph{lost} are the discordant counts $b$/$c$, prompts that moved to refusal and away from it, so each $\Delta$ is $(b-c)/100$ and each $p$ is an exact two-sided McNemar on that pair alone; both are recomputable from the two count columns. With mention held constant, attachment still moves refusal $+20$ points with \emph{no} prompt moving the other way: the cue is not a lexical artifact. \textbf{The mention cell is underpowered rather than null}: at $16$ discordant pairs the exact test could only have resolved $\sim$$10$ points (\S\ref{sec:res-threshold}).}
\label{tab:factorial}
\end{table}

\label{app:support}

Each table here is printed in support of a claim the main paper states with its numbers in prose; a reader checking a number will want the table, a reader evaluating the claim does not need it.

\begin{table}[t]
\centering\small
\begin{tabular}{lccc}
\toprule
 & \multicolumn{3}{c}{$\Delta$ benign refusal, text $\to$ blank (points)} \\
\cmidrule(lr){2-4}
OR-Bench category & claude & 4o-mini & fl.-lite \\
\midrule
deception   & \textbf{+80} & \textbf{+43} & $+13$ \\
harassment  & \textbf{+53} & \textbf{+23} & \textbf{+27} \\
harmful     & \textbf{+47} & \textbf{+53} & $+20$ \\
hate        & \textbf{+57} & \textbf{+33} & \textbf{+53} \\
illegal     & \textbf{+53} & \textbf{+33} & \textbf{+43} \\
privacy     & \textbf{+50} & \textbf{+43} & \textbf{+40} \\
self-harm   & \textbf{+63} & \textbf{+27} & \textbf{+37} \\
sexual      & \textbf{+40} & \textbf{+27} & \textbf{+57} \\
unethical   & $+20$        & \textbf{+43} & $+20$ \\
violence    & \textbf{+60} & \textbf{+37} & \textbf{+37} \\
\midrule
\emph{all $300$}          & \textbf{+52.3} & \textbf{+36.3} & \textbf{+34.7} \\
\emph{text baseline} (\%) & \emph{9.3}     & \emph{19.0}    & \emph{24.0} \\
\bottomrule
\end{tabular}
\caption{[\textsc{primary test}]~\textbf{The benign cost, stratified across all ten OR-Bench categories.}
$30$ prompts per category, $300$ per model, borderline-benign rung, no defense, both
arms collected in one job per model, paired per prompt id. \textbf{Bold}: survives
Benjamini--Hochberg across all $30$ per-category tests ($26$ of $30$ do; the four that
do not are claude \emph{unethical} $p=0.070$ and flash-lite \emph{deception}
$p=0.219$, \emph{harmful} $p=0.070$, \emph{unethical} $p=0.070$). Every contrast is
positive. Aggregates: claude $158$ discordant to $1$, $p=4.4\times10^{-46}$;
\texttt{gpt-4o-mini} $109$ to $0$, $p=3.1\times10^{-33}$; \texttt{fl.-lite} $110$ to
$6$, $p=7.6\times10^{-26}$. \texttt{gemini-2.5-flash} is not re-run here: it is the
null model on this rung, so we report it as untested on the stratified set rather
than implying it was checked. \textbf{One pre-specified check did not pass:} we
predicted claude's no-image baseline would sit above its $12\%$ on the original
slice, and it came in at $9.3\%$, so the strata differ in base sensitivity for
that model and the per-category deltas, not the baseline, are the quantity to read
(\S\ref{sec:res-threshold}).}
\label{tab:strata}
\end{table}

\begin{table}[t]
\centering
\small
\setlength{\tabcolsep}{5pt}
\begin{tabular}{lrrr}
\toprule
arm & managed & self-served & $\Delta$ \\
\midrule
text (no image)         & $6$  & $5$  & $+1$ \\
blank $1024\times768$   & $13$ & $12$ & $+1$ \\
blank $1024\times141$   & $13$ & $12$ & $+1$ \\
caption $1024\times141$ & $6$  & $9$  & $-3$ \\
\bottomrule
\end{tabular}
\caption{[\textsc{exploratory follow-up}]~\textbf{Same weights, two serving stacks.} Benign refusal rate (\%) for \texttt{gemma-3-12b-it}, the same open checkpoint reached through a managed commercial host and through our own vLLM, on the borderline-benign rung with no defense, $100$ prompts, one canonical text, paired per prompt id. Route differences are bounded within $\pm10$ points on every arm and within $\pm5$ points on both blank arms ($95\%$ Newcombe intervals; $1$--$3$ discordant pairs of $100$; exact McNemar $p\geq0.25$), while a blank canvas inflates refusal $+7$ points on both. Weights, vision encoder, pretraining and post-training alignment are identical across the two columns by construction, so the comparison isolates the inference implementation and vendor wrapper, and finds them inert (\S\ref{sec:res-threshold}).}
\label{tab:sameweights}
\end{table}

\begin{table}[t]
\centering
\scriptsize
\setlength{\tabcolsep}{3pt}
\begin{tabular}{lccccc}
\toprule
Rung & text & blank & $\Delta$ & disc. & $p$ \\
\midrule
neutral                       & \phantom{0}2 & \phantom{0}5 & +3 & 4/1 & 0.375 \\
\textbf{borderline}           & 53 & \textbf{82} & \textbf{+29} & 30/1 & $\mathbf{<10^{-4}}$ \\
harmful \emph{(ASR)}          & \phantom{0}2 & \phantom{0}1 & $-1$ & 1/2 & 1.000 \\
\bottomrule
\end{tabular}
\caption{[\textsc{replication}]~\textbf{Both sides of the price on one open checkpoint} (\texttt{qwen3-vl-8b}), all cells in one job, $100$ prompts each, no defense. Benign rungs are refusal rate (JailbreakBench refusal classifier); the harmful rung is ASR (HarmBench classifier) on OR-Bench's harmful split, the matched counterpart of the borderline rung. The ladder structure replicates on a downloadable model. \textbf{The harmful contrast is an underpowered null, not a zero}: this checkpoint already yields a harmful completion on only $2\%$ of plain harmful requests, so there is almost nothing left for the cue to prevent, and we do not claim it buys nothing. With the floor removed by attack the cue does measurably act on this checkpoint, in both directions depending on encoding (Table~\ref{tab:headroom}).}
\label{tab:owladder}
\end{table}

\begin{table}[t]
\centering
\scriptsize
\setlength{\tabcolsep}{3pt}
\begin{tabular}{lccrrr}
\toprule
encoding & no image & $+$canvas & $\Delta$ & gain/loss & $p$ \\
\midrule
plain            & 17 & \phantom{0}7 & $\mathbf{-10}$ & 0/10 & $\mathbf{0.0020}$ \\
set theory       & 52 & 40 & $\mathbf{-12}$ & 7/19 & $\mathbf{0.029}$ \\
formal logic     & 51 & 46 & $-5$           & 11/16 & 0.442 \\
code             & 63 & 73 & $+10$          & 16/\phantom{0}6 & 0.052 \\
\bottomrule
\end{tabular}
\caption{[\textsc{primary test, pre-specified}]~\textbf{The harmful side with the floor removed} (\texttt{qwen3-vl-8b}, $100$ HarmBench behaviours per cell, no defense, one job). Cells are attack success rate under the HarmBench classifier; $p$ is an exact paired McNemar test over the same $100$ behaviours, and gain/loss counts the discordant pairs in each direction. The three non-plain encodings are established published attacks used here as instruments and not proposed by us: set theory and formal logic follow the mathematical-encoding attacks of \citet{zhang2026exposingllmsafetygaps}, and the code row follows CodeAttack \citep{ren2024codeattack}. Encoded text is byte-identical across each row's two arms, so the $512^2$ white canvas is the only difference. Every no-image arm clears the $10$--$90\%$ band we fixed in advance for a contrast to count as powered. \textbf{The cue is not inert here}, but it helps on three arms and hurts on one, and which it does depends on how the request is written (\S\ref{sec:res-threshold}).}
\label{tab:headroom}
\end{table}

\begin{table*}[t]
\centering\small
\begin{tabular}{p{0.42\textwidth}p{0.46\textwidth}}
\toprule
Claim & Status \\
\midrule
\multicolumn{2}{l}{\emph{\textbf{Measured}: direct paired contrasts, reported with intervals}} \\[2pt]
Image presence raises benign refusal on particular aligned checkpoints ($+23$ to $+54$ points, hosted; $+28$ points on an open one) & measured \\
$\star$~The cost is spread across all ten OR-Bench categories, not two & measured (Tab.~\ref{tab:strata}) \\
$\star$~The price varies with canvas colour and pixel count & measured (Tab.~\ref{tab:imgvsimg}), with an image attached in \emph{both} arms of every contrast; colour on the hosted models rests on one file per colour, which is structural, since flat fills admit no independent renders \\
$\star$~The pixel-count price is not an input-length effect & measured (Tab.~\ref{tab:tokenbudget}) \\
The sign inverts on two open checkpoints ($+33$, $+39$ points ASR) & measured \\
An \emph{asserted} attachment moves refusal with no image attached ($+16$ points over a matched system message) & measured (Tab.~\ref{tab:placebo}) \\
On the models where we measure both sides, the benign cost does not track the harmful benefit & measured \\[4pt]
\multicolumn{2}{l}{\emph{\textbf{Control-supported}: rests on a control with a stated bound}} \\[2pt]
The effect can arise with no hosting layer at all & self-served checkpoint at $+32$ points; route bound $\pm10$ points on a $+7$--$9$ points checkpoint only; the hosted stacks' own contribution \emph{not} identified \\
The decoupling is not an artifact of comparing unlike models & within-family ladder, all rungs bounded $\pm10$ points \\
The behaviour steps between two adjacent checkpoints of one family & three rungs, one family, one size class \\
The judge does not manufacture the effect & survives a cross-family judge; both rubrics human-anchored, $\kappa=0.79$/$0.68$; differential $+3.8$ points \\
$\star$~The cue is not lexical: attachment moves refusal with the mention held byte-identical & $+20$ points, $20$ gained / $0$ lost (Tab.~\ref{tab:factorial}) \\[4pt]
\multicolumn{2}{l}{\emph{\textbf{Hypothesis}: consistent with our data, not established by it}} \\[2pt]
Post-training specifically (rather than vision encoder or corpus) introduces the behaviour & \emph{not identified}: releases change all three \\
The harmful-side denominator shrinks with model recency & across-model only; our family is floored throughout \\
Attachment carries no risk information in real traffic & \emph{not claimed}: unmeasurable black-box \\
Which property of the visual input sets the magnitude & \emph{not identified}: candidates excluded, the positive question left open (\S\ref{sec:res-notwhat}) \\
\bottomrule
\end{tabular}
\caption{\textbf{What this paper measures, what it infers from a control, and what it
does not establish.} Tier 3 is deliberately non-empty: each row is a reading a reader
might otherwise take from the prose, listed with the reason our design cannot support it. $\star$ marks the four rows carrying the most evidential weight: the category-stratified benign cost, the colour and pixel-count contrasts taken with an image attached in both arms, the token-matched control, and the mention-held factorial. A reader who reads only those four rows has this paper's load-bearing evidence; every other row supports, bounds or qualifies them.}
\label{tab:claims}
\end{table*}

\section{Interpretability findings that bear on the result}
\label{app:interp}

Two findings from the interpretability literature bear on why a channel carrying nothing could matter at all. We cite them as hypotheses rather than explanations, since neither studies safety. \citet{kim2025visualabsence} show that vision--language models routinely treat a textual concept with no visual evidence behind it as though the image contained it, and locate feed-forward neurons whose activation signals that absence: a mechanism whose shape matches our placebo result, where asserting an attachment moves refusal as much as attaching one does elsewhere. \citet{zhou2026visualignorance} run the complementary diagnosis and find visual information retrieved weakly in intermediate layers and further suppressed in later ones, so that answers across twelve visual question-answering benchmarks often survive severe or total visual obfuscation. Set beside our result the picture is an awkward one: the image's \emph{content} is frequently ignored, and yet its \emph{presence} moves the safety threshold by tens of points. Whether one pathway carries both is the question a white-box study could now settle, and \texttt{qwen3-vl-8b} is a downloadable place to ask it.

\section{Limitations, in full}

\label{app:limits}

The main paper states each limitation; this is the elaborated form.

\paragraph{Scope of the claim.}
Table~\ref{tab:claims} sorts every claim in this paper by what supports it (a direct
paired measurement, a control with a stated bound, or neither), and the limitations
below elaborate the entries that need it.
We do not claim the image-presence effect is universal across VLM architectures; we describe a pattern measured on four frontier hosted models and seven open-weight checkpoints, and we report the models on which it is absent as prominently as those on which it is large. Three of the seven open-weight models show no benign cost at all, and a sample of that kind invites a claim about model classes, which the fourth open checkpoint then refuted (\S\ref{sec:res-threshold}). We therefore state the positive claim narrowly: the behaviour is a property of \emph{particular aligned checkpoints}, hosted or open, and we cannot predict from any label we have tried which checkpoints will have it. The sign reversal on \texttt{pixtral-12b} rests on one model and establishes that the direction is not fixed, not what fixes it. Being black-box throughout, we locate the effect in the checkpoint rather than the serving layer, but we do not identify what inside the checkpoint produces it, and in particular we do not isolate post-training from the vision encoder or pretraining corpus that a release changes alongside it.

\paragraph{Model coverage.}
The full ladder plus matched harmful set is complete on four hosted models and, on the open side, on the one checkpoint that carries the scope claim; the other six open checkpoints are measured on the borderline rung only. The generational ladder is one family of three rungs at one size class, so \emph{when} the behaviour appears is established for that family and not in general; and because that family is already floored on the harmful split at its oldest rung ($2\%$), it can demonstrate the cost--benefit decoupling under control but cannot test the \emph{trend} in harmful headroom, which stays an across-model observation. Our manipulation also varies the attachment interface as a bundle on the hosted models (serialisation, modality tokens, vision-encoder routing and provider preprocessing move together whenever an image is attached), so there we report an interface-level effect and do not separate those four. On \texttt{qwen3-vl-8b} we separate them in part, excluding provider preprocessing, total input length, serialisation and encoder invocation as explanations of a magnitude that varies across arms. Which property of the visual input then carries that magnitude is left open here, and it is the one question about this effect that a black-box design on the hosted tier cannot even pose. The factorial below separates attachment from \emph{mention}, which is a different cut. The attachment$\times$mention factorial is run on one checkpoint; the hosted model whose behaviour motivated it remains unresolved.

\paragraph{The serving-route control is run on a small-effect checkpoint.}
The same-weights comparison holds weights and post-training fixed by construction, but it
is run on \texttt{gemma-3-12b-it}, whose own presence effect is $+7$ to $+9$ points. Our route
bound ($\pm10$ points on every arm, $\pm5$ points on both blank arms) therefore sits at roughly the
scale of the effect it is meant to validate, and we do not claim it settles the route
question at that model's effect size: only that it excludes a route artifact as the
source of the $+23$ to $+54$-point shifts we report on the hosted models. The stronger design
is obvious: route-test a checkpoint with a \emph{large} effect, such as
\texttt{qwen3-vl-8b} at $+28$ points. We did not, because it is not currently possible. That
test needs a checkpoint that is simultaneously vision-capable, offered by a managed
commercial host, and small enough to self-serve on the GPUs we have. On the managed host
available to us the vision-capable open-weight families are \texttt{gemma-3},
\texttt{llama-4}, \texttt{pixtral-large} and \texttt{qwen3-vl}, and only \texttt{gemma-3}
satisfies the third condition: \texttt{qwen3-vl} is offered at $235$B, which we cannot
self-serve. The checkpoint whose effect we would most want to route-test is thus
unavailable in the one configuration that would test it. We flag this as a named gap
rather than a resolved question.

\paragraph{Image coverage, and between-image variance.}
The attached files we test are a blank canvas at five sizes and three colours, two file encodings, a clip-art line drawing and a rendered caption. Each property level is represented by a \emph{single} fixed file repeated across the $100$ prompts, which gives paired prompt-level power for that file but does not estimate variance \emph{between} images within a property class, so the colour and content effects we report could in principle be idiosyncratic to the particular rendered artifacts. We therefore report them as paired contrasts between two fixed files rather than as property-class estimates. The claim they support, that the response is not invariant to variation carrying no information about the request, does not require the stronger reading, since a dependence on \emph{this} black canvas versus \emph{this} white one is already a dependence on something the request does not determine.

\paragraph{Judge calibration and benign benchmark.}
ASR is assigned by an LLM judge. Our claims are paired within-prompt contrasts scored by a held-constant instrument (\S\ref{sec:res-judge}), which is the property they depend on. Three judges have now been applied to these cells (\texttt{gpt-5-mini} at collection, \texttt{gpt-5-nano} across all $88$, and \texttt{gemini-2.5-pro} across the load-bearing ones) and every effect keeps its sign, significance and approximate magnitude; what remains untested is a blind spot common to all rubric-following language models, which only the two human anchors speak to (one per rubric, $100$ blind labels each) and which they bound only at $n\approx50$ per arm. Absolute ASR levels are separately judge-relative: judges differ by a factor of two to four on the same stored responses, disagreeing chiefly on whether partially-completed code artifacts count as successful attacks. Absolute numbers here should therefore not be compared against papers using a different judge. Our benign rungs are drawn from three sources, one per rung (AlpacaEval, OR-Bench, JailbreakBench-benign), and the borderline rung that carries the headline cost comes from one of them; benign traffic that is sensitive in other ways (mathematical, code-heavy or multilingual text) is untested, so the population the cost falls on is characterised only as far as those sets reach.

\section{Why the route control is not run on the large-effect checkpoint}
\label{app:routecontrol}

A larger-effect version of this control is available in principle, and we say why we do not report it as one. Our open checkpoint is also served by managed hosts, so running the identical manipulation through one would bound the route contribution against a $+28$-point effect rather than gemma's $+7$ points, which is the natural objection to the arm above. The obstacle is not access but identification. The magnitude on that checkpoint tracks the number of visual tokens the canvas becomes, and how many visual tokens a given canvas becomes is decided by the serving stack's image preprocessing: the resize bounds a vLLM deployment applies before patching are configuration, not a property of the weights, and a managed endpoint does not expose them. A route comparison run naively therefore varies the input the model actually receives at the same time as the route, and a difference between the two columns would be unattributable. The design requirement that would fix it is stated easily enough, and we record it for anyone running this control, ourselves included: read back the host's own prompt-token accounting per arm and confirm the canvas tokenises to the same count on both routes before comparing refusal. The gemma arm above is not exposed to this, because its contrast is carried by one to three discordant pairs and is reported as a bound rather than as an estimate.

\section{Where our outcome measure is coarser than the literature's}
\label{app:outcome}

That a visual input can move safety behaviour in either direction is also what \citet{ren2026seeingthreat} report from the adversarial side, where refusal, instruction non-compliance and successful exploitation appear as distinct outcomes of one manipulation; their response taxonomy (direct refusal, soft refusal, and refusal that is nonetheless helpful) is a finer instrument than the binary refusal classifier we use, and partial compliance of that kind is scored here by the HarmBench classifier on the completion rather than split out as its own category. A binary refuse/comply outcome also collapses a category that a finer taxonomy separates: \citet{yang2026refusalreframing} distinguish outright refusal from \emph{reframing}, where a model answers but redirects the framing of the answer. Our judges score refusal and harm, not reframing, so a shift from answering to reframing under attachment would register here only if it crossed one of those two lines: a narrower outcome than the behaviour space actually contains.

\section{Collection windows and the byte-identical design}
\label{app:windows}

\paragraph{Why every arm is byte-identical within itself.}
A referee might ask for images randomised \emph{across} prompts. We deliberately keep every arm byte-identical internally, because that is what licenses the paper's central claim that the cue carries no per-prompt information; randomising files across prompts would trade that control away. Varying the instance \emph{between} arms, as the contrasts above do, estimates between-image variance without giving up the design's main virtue.

\paragraph{The rates replicate across collection windows.}
This job re-collected the blank-canvas arm from scratch, giving an unplanned independent replication of the ladder's rung~2: \texttt{claude} $12\!\to\!63$ vs $10\!\to\!64$, \texttt{gpt-4o-mini} $12\!\to\!46$ vs $13\!\to\!43$, \texttt{gemini-2.5-flash-lite} $11\!\to\!34$ vs $11\!\to\!34$, \texttt{gemini-2.5-flash} $16\!\to\!13$ vs $15\!\to\!14$. This matters because these targets are \emph{not} deterministic at temperature $0$: the same checkpoint's text baseline reads $51$, $54$ and $53\%$ across the three open-weight jobs below. Response-level nondeterminism coexists with cell-level rates stable to a few points, which is why every contrast in this paper is collected within a single job and why the aggregate quantities are nonetheless reproducible.

\section{Arm scheduling, in full}
\label{app:scheduling}

Because we collect arms through live APIs and observe response-level nondeterminism at temperature zero, the temporal arrangement of arms is part of the design and we state it exactly. Arms within a contrast are dispatched as concurrent tasks of a single job under one shared concurrency limit, so both arms of a pair are issued together and their collection windows overlap rather than one arm running to completion before the other begins. Across the $127$ paired text/image arms in our collection records, \textbf{every pair starts within the same second}, and the two windows overlap by a median of $77\%$ of their combined span (mean $73\%$, range $10$--$100\%$). Batching, concurrency and retry logic are set by the same code path for both arms and do not vary by condition. We are explicit about what this does \emph{not} include: prompts are batched per arm rather than interleaved across arms, and arm order within the dispatch is not randomised. Concurrent dispatch makes a clean ordering confound unlikely, a provider-state drift would have to fall inside overlapping windows and align with condition rather than with time, but it does not exclude one by construction the way randomised paired interleaving would. We therefore ran that design directly, as a control on this one.

\section{The randomised-interleaving control}
\label{app:interleave}

To remove the scheduling confound by construction rather than bound it, we re-collected the central borderline contrast under a design in which the two conditions are never separately dispatched. For each prompt we emit one request per condition, shuffle the combined $200$ requests under a recorded seed, and issue them as a \emph{single} batch, so any transient serving state is shared across conditions instead of varying with them. Both arms are then scored by the same pinned judge and compared with the same paired exact test. We ran three targets under three independent shuffles. Each interleaved contrast is compared against the arm-batched contrast computed on the \emph{identical} upstream image file rather than against a per-model average, since magnitude depends on which content-free canvas is used. The two designs agree on all three targets: $+18.0$ against $+22.0$ points on \texttt{qwen3-vl-8b}, $+23.0$ against $+23.0$ on \texttt{gemini-2.5-flash-lite}, and $+42.0$ against $+38.0$ on \texttt{claude-sonnet-4-6}. The differences are $-4.0$, $0.0$ and $+4.0$ points, they do not share a sign, and each interleaved interval contains its matched arm-batched estimate. Two caveats belong with this result. First, the agreement is exact on \texttt{gemini-2.5-flash-lite} because that model returned byte-identical responses to the earlier collection on all $200$ requests; nothing varied between the designs there, so that target cannot evidence the absence of a batching artifact, only the absence of anything to detect. The two targets that are genuinely nondeterministic under repetition, \texttt{claude-sonnet-4-6} and \texttt{qwen3-vl-8b}, are the ones that carry this control. Second, we report interleaving as a control and retain the batched collection as the primary design, since each arm being a single homogeneous batch is what the rest of our provenance checks are built on.

\section{Delimitation from the closest prior work, in detail}
\label{app:zoudetail}

The closest prior work is \citet{zou2026understanding}, who study how image inputs distort VLM safety perception and use the same blank-image control we do, arguing on that basis that the shift originates in the visual modality itself rather than in image content: our design choice is theirs, and our two-image control extends the same argument. Two differences matter. Their analysis is directed at the harmful axis and at the direction in which distortion \emph{weakens} safety, which is the opposite direction from the tightening we measure on frontier hosted models. Our open-weight arm resolves that as a genuine sign difference rather than a disagreement: under one manipulation, image presence tightens the threshold on four moderated hosted models and loosens it $33$ points on \texttt{pixtral-12b} (\S\ref{sec:res-threshold}). Their finding and ours are the two signs of one effect, and the split is \emph{not} along the hosted/open line: \texttt{pixtral-12b} and \texttt{qwen3-vl-8b} are both open checkpoints we serve ourselves under one arrangement, and they move in opposite directions. A result established on one VLM should therefore not be assumed to transfer to another, hosted or open. More importantly, their benign-side table reports the effect of \emph{their proposed correction} on false-alarm rates rather than the effect of image presence itself, so the cost side (what image presence does to benign refusal, set against what it prevents) is not quantified there. That gap is the quantity this paper reports.

\section{Delimitation from concurrent work}
\label{app:novelty}

\citet{zou2026understanding} share our central design choice, a request-independent blank image as the control, and reach it independently. The differences are not framing: they are the benign-cost accounting, the matched harmful counterpart, the property and placebo controls, and the open-checkpoint and serving-route arms. Table~\ref{tab:novelty} sets the two out side by side; the main paper's discussion gives the full comparison, including where our sign result and theirs disagree and why we read them as two signs of one effect.

\begin{table}[t]
\centering
\scriptsize
\setlength{\tabcolsep}{3pt}
\begin{tabular}{p{0.44\columnwidth}p{0.20\columnwidth}p{0.24\columnwidth}}
\toprule
 & \citet{zou2026understanding} & this paper \\
\midrule
request-independent blank image as the control & yes & yes \\
axis measured & harmful, weakening direction & benign \emph{and} harmful, both signs \\
benign refusal attributed to attachment itself & effect of \emph{their correction} on false alarms & effect of attachment, paired and stratified \\
matched harmful counterpart to the benign rung & --- & yes \\
image-property contrasts & --- & paired between-arm contrasts, image attached in both arms \\
mention placebo with \emph{no} image attached & --- & yes \\
open checkpoint served by us, no moderation layer & --- & yes \\
same-weights serving-route control & --- & bounded $\pm10$ points \\
\bottomrule
\end{tabular}
\caption{\textbf{What is new here relative to the closest prior work.} \citet{zou2026understanding} share our central design choice, a request-independent blank image as the control, and reach it independently; we state the delimitation here rather than after the related-work section so a reader can locate the contribution immediately. The differences are not framing: they are the benign-cost accounting, the matched harmful counterpart, the property and placebo controls, and the open-checkpoint and serving-route arms. \S\ref{sec:discussion} gives the full comparison, including where our sign result and theirs disagree and why we read them as two signs of one effect.}
\label{tab:novelty}
\end{table}

\section{Judge robustness: validating the instrument}
\label{sec:res-judge}

Refusal and harm are assigned by LLM judges, so the instrument deserves a statement. Every number in this paper was scored at collection time by \texttt{gpt-5-mini} running the standardized JailbreakBench refusal classifier (benign rungs) or the HarmBench classifier (harmful arms); no result here rests on a judge migration, and the two rubrics are never substituted for one another. What protects the claims is the design rather than the judge: every quantity we report is a \emph{paired within-prompt contrast} scored by a held-constant instrument, so a judge that is systematically strict or lenient shifts both arms together and leaves the contrast intact. Absolute levels are another matter: they should be read as judge-relative and not compared against papers using a different backbone.

That design argument covers a judge that is \emph{uniformly} strict, but not the sharper objection: attachment might change a response's \emph{style} (hedging, verbosity, disclaimers, partial compliance) and a judge could label the same underlying behaviour differently in the two arms, manufacturing a contrast where no threshold moved. Systematic strictness cancels in a paired design; differential error does not. We test it directly. Re-scoring all $88$ cells behind Tables~\ref{tab:ladder}--\ref{tab:ow_threshold} with a second judge (\texttt{gpt-5-nano}, same rubrics, applied to the \emph{stored} responses so no model is re-queried) leaves the paper's claims intact: \textbf{every contrast that is significant under \texttt{gpt-5-mini} keeps its sign and its significance under \texttt{gpt-5-nano}}, and the point estimates move by at most a couple of points: the borderline rung reads $+51$ and $+34$ points on \texttt{claude-sonnet-4-6} and \texttt{gpt-4o-mini} under both judges, $+23$ versus $+22$ points on \texttt{gemini-2.5-flash-lite}, and the harmful-side reduction on \texttt{claude-sonnet-4-6} is $-18$ points under both. The nulls stay null. Exactly one contrast changes status, and it is the one we already tell the reader not to trust: the \texttt{gemini-2.5-flash} line-drawing cell of Table~\ref{tab:presence}, $+10$ points at $p=0.021$ under \texttt{gpt-5-mini}, falls to $+8$ points at $p=0.096$ under \texttt{gpt-5-nano}: a third independent indication, after the property sweep's $+9$ points at $p=0.064$, that this cell is noise.

Both judges are OpenAI models, so that check bounds differential error \emph{within} a family and cannot exclude a blind spot the two share. We therefore re-scored the three results most load-bearing for the argument (the stratified benign cost of Table~\ref{tab:strata} on all three closed models, the low-headroom \texttt{qwen3-vl-8b} harmful null, and the \texttt{pixtral-12b} sign inversion of Table~\ref{tab:ow_threshold}) with \texttt{gemini-2.5-pro}, a judge from a different family applying the same two rubrics to the same stored responses ($2{,}200$ rescored rows; no model re-queried). The choice is deliberate rather than convenient: in our own judge-selection study over $17$ API rubric-appliers, \texttt{gemini-2.5-pro} is the strongest non-OpenAI candidate (the cheaper cross-family alternatives agree less with the human anchor and drift toward calling nothing harmful, which would floor the denominator and manufacture a null) and it is \emph{stricter} than \texttt{gpt-5-mini} on the harm rubric ($33\%$ versus $50\%$ flagged against that anchor), so it is the conservative instrument for a claim like ours. \textbf{Every effect survives.} On the stratified benign rung the paired shift is $+57.0$, $+34.0$ and $+34.7$ points for \texttt{claude-sonnet-4-6}, \texttt{gpt-4o-mini} and \texttt{gemini-2.5-flash-lite} ($172$ discordant to $1$, $108$ to $6$, $111$ to $7$; all $p<10^{-24}$), against $+52.3$, $+36.3$ and $+34.7$ points under \texttt{gpt-5-mini}. The cross-family judge sits at a visibly different absolute level (it calls $2$ to $12$ points more refusal on every benign cell, in both arms) and the paired contrast nonetheless lands within a few points everywhere, which is precisely the cancellation the paired design predicts and the reason absolute leniency was never the quantity at issue. The \texttt{pixtral-12b} inversion reads $+30.0$ points ($95\%$ CI $[+17.9,+40.9]$, $37$ to $7$, $p=5.3\times10^{-6}$) against $+33$ points. Raw verdict agreement between the two judges is $94.4\%$ across all $2{,}200$ rows. That number is \emph{agreement}, not accuracy: two judges concurring bounds instrument-specific error, and cannot establish that either matches ground truth. Two rubric-followers can agree and both be wrong. Only the human anchors below speak to correctness, and we keep the two quantities separate throughout. The \texttt{qwen3-vl-8b} harmful null stays null ($5\%$ versus $3\%$, $p=0.63$) and stays uninformative for the reason we already give: the stricter judge moves that cell to $5\%$, not into a range where a threshold shift could become visible.

Two caveats we would rather state than have found. First, \texttt{gemini-2.5-flash-lite} is itself one of the targets being scored here, so on that one row the judge shares a model family with the model it judges; we report per-model rather than pooled so a reader can discount it, and note that it is the contrast that does not move at all ($+34.7$ points under both judges). Second, the instrument has a noise floor of its own: re-running \texttt{gpt-5-mini} over the \emph{identical} stored \texttt{pixtral-12b} responses flips $2$ of $100$ verdicts, a $\pm2$ points spread on that cell ($+35$ versus $+33$ points); we carry the lower value throughout. Judge agreement is also weakest exactly where it means least: the two floored \texttt{qwen3-vl-8b} cells give $\kappa=0.49$ and $0.56$ on $97$--$98\%$ raw agreement, the familiar instability of $\kappa$ at a $1$--$5\%$ base rate, while the six benign cells run $\kappa=0.65$ to $0.94$.

A different-family judge still cannot exclude a blind spot shared by all rubric-following language models. For that we go to human labels. Our human anchor for the refusal rubric is $100$ benign responses labelled blind with the judge verdict withheld ($\kappa=0.79$ on the answered-versus-refused collapse the reported quantity is actually computed from; $\kappa=0.85$ under one-response-per-question resampling), and its sample is split $49$ text-channel against $51$ image-channel: an attachment contrast, though not this paper's blank-canvas one (scope note below). That lets us measure the arm-differential quantity the objection is really about, on the nearest manipulation we have labelled. Against the human labels the judge over-calls refusal by $+4.1$ points in the text arm and $+7.8$ points in the image arm; the component that does \emph{not} cancel in a paired contrast is the difference between those, $\mathbf{+3.8}$\textbf{ points} ($95\%$ CI $[-8.3,+15.9]$, $10$k row-resamples within arm). Agreement is if anything \emph{higher} in the image arm ($\kappa=0.84$ versus $0.74$): the opposite of what a judge confused by attachment-induced style would produce.

The harm rubric has its own human anchor, and it is worth stating separately because the refusal anchor says nothing about it. A second blind round of $100$ responses (hand-labelled with the judge verdict withheld, under the same HarmBench criteria the automatic judge applies) puts \texttt{gpt-5-mini} at $\kappa=0.68$ against the human labels on the harm task. Its sample is split $56$ text-channel against $44$ image-channel, and $32$ of its $100$ rows are \texttt{pixtral-12b}, the model carrying the sign inversion this section is asked to defend. So both rubrics behind this paper's two axes are humanly calibrated, and the harm side is calibrated on one of the two models whose harmful contrast we report.

We decline to round that into ``the bias cancels''. The interval contains zero, and its upper end is $29\%$ of our largest effect but $69\%$ of our smallest significant one, so this test protects the contrasts of $+34$ to $+54$ points and does \emph{not} fully protect the $\sim\!23$ points ones; $n\approx50$ per arm is the binding constraint. One scope note belongs with it: those labels come from a defense-pipeline study of ours on self-served targets, where the image channel carried a payload rather than a blank canvas, so they bound the judge's arm-dependent bias on a neighbouring manipulation rather than on these exact cells. Every per-prompt row stores the full judge audit trail (\texttt{judge\_output}, \texttt{judge\_reasoning}, \texttt{judge\_raw\_response}) alongside the response, so any individual judgement can be inspected without re-running anything.

\end{document}